\documentclass{aastex}
\usepackage{emulateapj5}

\usepackage{epsfig}
\usepackage{rotate}

\def \ie{{\it i.e.}}
\def \eg{{\it e.g.}}
\def \kid{$\chi^2$}
\def \deltakid{$\Delta\chi^2$}

\def \lya{Ly-$\alpha$}
\def \lyb{Ly-$\beta$}

\def \cmmd{cm$^{-2}$}

\def \dshism{(D/H)$_{\mathrm{ISM}}$}

\def \dshlism{(D/H)$_{\mathrm{LISM}}$}
\def \ow{{Owens.f}}

\def \fuse{{\it FUSE}}
\def \2s{$2\,\sigma$}
\def \1s{$1\,\sigma$}

\shorttitle{The D/O ratio in the interstellar medium}
\shortauthors{H\'ebrard \& Moos}

\begin{document}

\title{The deuterium-to-oxygen ratio in the interstellar medium}


\author{Guillaume H\'ebrard\altaffilmark{1}}
\affil{Institut d'Astrophysique de Paris, CNRS, 
       98$^{bis}$ boulevard Arago, F-75014 Paris, France}
%\email{hebrard@iap.fr}

\and

\author{H. Warren Moos}
\affil{Department of Physics and Astronomy, Johns Hopkins University, 
       Baltimore, MD 21218, USA}

\altaffiltext{1}{email: hebrard@iap.fr}

\begin{abstract}
Because the ionization balances for \ion{H}{1}, \ion{O}{1}, and
\ion{D}{1} are locked together by charge exchange, the
deuterium-to-oxygen ratio, D/O, is an important tracer for the value
of the D/H ratio and for potential spatial variations in the ratio.
As the \ion{D}{1} and \ion{O}{1} column densities are of similar
orders of magnitude for a given sight line, comparisons of the two
values will generally be less subject to systematic errors than
comparisons of \ion{D}{1} and \ion{H}{1}, which differ by about five 
orders of magnitude.
Moreover, D/O is additionally sensitive to astration, because as stars
destroy deuterium, they should produce oxygen.
We report here the results of a survey of D/O in the
interstellar medium performed with the Far Ultraviolet Spectroscopic
Explorer (\fuse). We also compare these results with those for D/N.
Together with a few results from previous missions, the sample totals
24 lines of sight. The distances range from a few pc to $\sim2000$~pc
and $\log N$(\ion{D}{1}) from $\sim13$ to $\sim16$~(\cmmd). The D/O
ratio is constant in the local interstellar medium out to distances of
$\sim150$~pc and $N$(\ion{D}{1})$\,\simeq 1 \times 10^{15}\,$\cmmd, 
\ie\ within the Local Bubble. In this region of the interstellar
space, we find D/O$\;= (3.84\pm0.16) \times 10^{-2}$ ($1\,\sigma$ in
the mean).  The homogeneity of the local D/O measurements shows that
the spatial variations in the local D/H and O/H must be extremely
small, if any.  A comparison of the Local Bubble mean value with the
few D/O measurements available for low metallicity quasar sight lines
shows that the D/O ratio decreases with cosmic evolution, as
expected. Beyond the Local Bubble we detected significant spatial
variations in the value of D/O. This likely implies a variation in
D/H, as O/H is known to not vary significantly over the distances
covered in this study. Our dataset suggests a present-epoch deuterium
abundance below $1\times 10^{-5}$, \ie\ lower than the value usually
assumed, around $1.5\times10^{-5}$.
\end{abstract}

%%The uncommented

\keywords{ISM: abundances -- ISM: clouds -- cosmology: observations -- 
ultraviolet: ISM -- (stars:) white dwarfs-- (stars:) subdwarfs}

\section{Introduction}

It is generally believed that deuterium was produced in significant
amounts only during the primordial nucleosynthesis of the Big Bang
(BBN).  Since then, deuterium has been steadily destroyed in stellar
interiors by nuclear processes. Thus, its abundance compared to
hydrogen, D/H, is a key measurement for studies of both cosmology and
galactic chemical evolution (see, \eg, Vangioni-Flam et
al.~\citealp{flam00}). Although the evolution of the deuterium
abundance seems to be qualitatively understood, important questions
remain open.

In this paper we focus on the abundance of deuterium in the
interstellar medium (ISM), which is characteristic of the present-day
Galactic deuterium abundance. \dshism\ measurements prior to the Far
Ultraviolet Spectroscopic Explorer (\fuse) mission showed some
dispersion (see, \eg, Lemoine et al.~\citealp{lemoine99} for a
review).  This dispersion likely results from poorly understood
physical processes (\eg\ astration and transport), although the
effects of unidentified systematic errors cannot be completely ruled
out.  There has been considerable debate of these issues and final
resolution will have implications for our understanding of the physics
of the interstellar medium, the chemical evolution of the Galaxy, and
the baryonic density of the Universe inferred from primordial D/H
measurements (see also Moos et al.~\citealp{moos02}; Hoopes et
al.~\citealp{hoopes03}).

An accurate determination of interstellar deuterium abundances is one
of the main objectives of the \fuse\ mission, which was launched in
1999 (Moos et al.~\citealp{moos00}).  The \fuse\ bandpass ranges from
905~\AA\ to 1187~\AA, so it includes all the \ion{D}{1} transitions of
the Lyman series except for \lya.  We report here the first results of
a survey of D/O in the ISM.  This survey includes already published
\fuse\ results, together with additional new \fuse\ results obtained
toward seven targets.  The first \fuse\ measurements of D/O were
presented by Moos et al.~(\citealp{moos02}) and references therein.
This earlier study was based on only seven targets and a good 
understanding of the D/O ratio was limited by low number
statistics.  Additional measurements by other authors and the new ones
reported here increase the number of \fuse\ sight lines to
19. Together with five measurements from previous missions, the
present study includes abundance measurements toward 24 lines of sight
in the Galactic disk.  The sample includes a wide variety of sight
lines; the deuterium column densities vary over three orders of
magnitude and the sight line distances range from within the Local
Bubble to $\sim2000$~pc.

The measured deuterium abundance by number is usually compared to that
of hydrogen.  One of the challenges of D/H measurements is to evaluate
two measurements for the same line of sight, namely the \ion{H}{1} and
\ion{D}{1} column densities, which differ by about five orders of
magnitude. Such a large difference is a potential source of systematic
errors. For example, all lines from the same species may lie on the
non-linear part of the curve of growth, or there may be clouds in
which \ion{H}{1} column densities are detectable, but whose \ion{D}{1}
column densities are below the detection limit (see, \eg, Linsky \& 
Wood~\citealp{linsky96}, Lemoine et al.~\citealp{lemoine02}, or
Vidal-Madjar \& Ferlet~\citealp{avm02}).

Many of the difficulties associated with obtaining accurate D/H
measurements may be avoided by measuring the deuterium-to-oxygen
ratio, D/O (see, \eg, Timmes et al.~\citealp{timmes97}). First of all,
as discussed above, the D/O ratio is of order of a few percent rather
than $\sim10^{-5}$ as for the case of D/H, thus the \ion{D}{1} and
\ion{O}{1} column density measurements are not subject to some of the
systematic errors associated with measurements of D/H.  In addition to
the \ion{D}{1} transitions, many \ion{O}{1} transitions with different
oscillator strengths are present in the \fuse\ bandpass, allowing
measurement of the \ion{O}{1} and the \ion{D}{1} column densities over
a wide range of values. \ion{O}{1} is believed to be a good tracer of
\ion{H}{1} in the nearby Galactic disk (Meyer et al.~\citealp{meyer98}; 
Andr\'e et al.~\citealp{andre03}). The neutral forms of oxygen and
hydrogen likely dominate over their ions for many sight lines in the
diffuse interstellar medium.  In any case, because both species have
nearly the same ionization potentials, their ionization balances for
the two species are strongly coupled to each other by charge exchange
reactions (Jenkins et al.~\citealp{jenkins00}). Thus, no corrections
from ionization models are required, and we will use D/O for
$N$(\ion{D}{1})/$N$(\ion{O}{1}) hereafter.  Finally, D/O is very
sensitive to stellar activity, because of both deuterium destruction
(deuterium is burned in stellar atmospheres at temperatures as low as
$6\times10^6$~K) and oxygen production (oxygen is mainly produced by
type~II supernovae).  Hence, spatial variations of the deuterium
abundance due to different astration rates in different locations
would cause even higher D/O spatial variations.  We use a D/O point of
view in the present paper, consciously avoiding a reliance on
\ion{H}{1} column density measurements.  We also compare D/O {\it vs.}
D/N, since nitrogen is often assumed to be also a good hydrogen tracer
(Ferlet~\citealp{ferlet81}; Meyer et al.~\citealp{meyer97}).

The new \fuse\ data and their processing are presented in
Sect.~\ref{Observations_and_data_processing}, and the method of
analysis in Sect.~\ref{Data_analysis}. The individual targets for
which we present new \fuse\ column density measurements here are
discussed in Sect.~\ref{Individual_targets}. The first results of 
the survey are reported in Sect.~\ref{Results} and discussed in
Sect.~\ref{Discussion}.

\section{Observations and data processing}
\label{Observations_and_data_processing}

The new \fuse\ data presented here were obtained in 2000, 2001, and
2002 as part of Science Team D/H programs, calibration programs, and a
Guest Investigator program. The log of the observations is reported in
Table~\ref{table_obslog}. They were obtained in histogram or
time-tagged modes, through the low (LWRS), medium (MDRS), or high
(HIRS) resolution slit. Details of the \fuse\ instrument may be found
in Moos et al.~(\citealp{moos00}) and 
Sahnow et al.~(\citealp{sahnow00}).

The one-dimensional spectra were extracted from the two-dimensional
detector images and calibrated using the CalFUSE pipeline. We used the
most recent version of CalFUSE available at the time of the
extraction.  The different improvements of CalFUSE include, among
others, corrections of event bursts, thermal drift, geometric
distortions, or astigmatism. No significant effects on column density
measurements were found between these different versions.  
In particular, no effects on the zero flux level were found.
Each \fuse\ 
observation is split up into individual exposures.  They were co-added
separately for each channel (SiC1, SiC2, LiF1, and LiF2) and for a
given slit, after correcting for %zero-point 
relative wavelength offsets between individual calibrated
exposures (typically, a few \fuse\ pixels). This correction procedure
improved the spectral resolution slightly and helped average out fixed
pattern noise in the detector, acting as a random FP-Split procedure
(Kruk et al.~\citealp{kruk02}).  Exposures with strong airglow
emissions or unacceptably low signal-to-noise ratios were not included
in the final sums.

All of the spectra were binned to three \fuse\ pixels.  The resolving
power in the final spectra ranges between $\sim13000$ and $\sim18500$
(full width at half maximum, \ie\ around 10 \fuse\ pixels), depending
on channel and wavelength, and on the slit used. These values are
similar to those found previously on \fuse\ data obtained toward
similar point-like sources and extracted with versions 1.7 or 1.6 of
CalFUSE (H\'ebrard et al.~\citealp{hebrard02}; 
Wood et al.~\citealp{wood02}).

\section{Data analysis}
\label{Data_analysis}

Column density measurements were obtained by profile fitting of the
interstellar spectral absorption lines obtained towards the lines of
sight studied. We focus here on results for the \ion{D}{1},
\ion{O}{1}, and \ion{N}{1} column densities, in order to compare D/O
and D/N results. We used the profile fitting method presented in 
detail by H\'ebrard et al.~(\citealp{hebrard02}). It is briefly 
described below.

\begin{figure}
\psfig{file=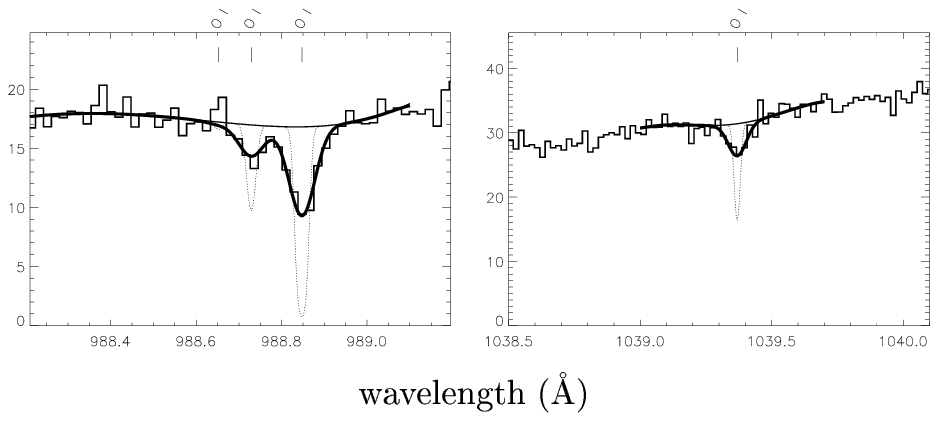,height=6.5cm}
\caption{Examples of \fuse\ spectral windows fitted on the line 
of sight of Sirius~B. 
The list of all the \ion{O}{1} 
lines included in the fits are reported in Table~\ref{table_lines}. 
Histogram lines are the data, the solid lines are the 
fits and continua, and the dotted lines are the model profiles 
prior to convolution with the LSF. 
Y-axis is flux in $10^{-12}\,$erg/cm$^2$/s/\AA.
\label{sirius_fit_plot}}
\end{figure}

\begin{figure}
\psfig{file=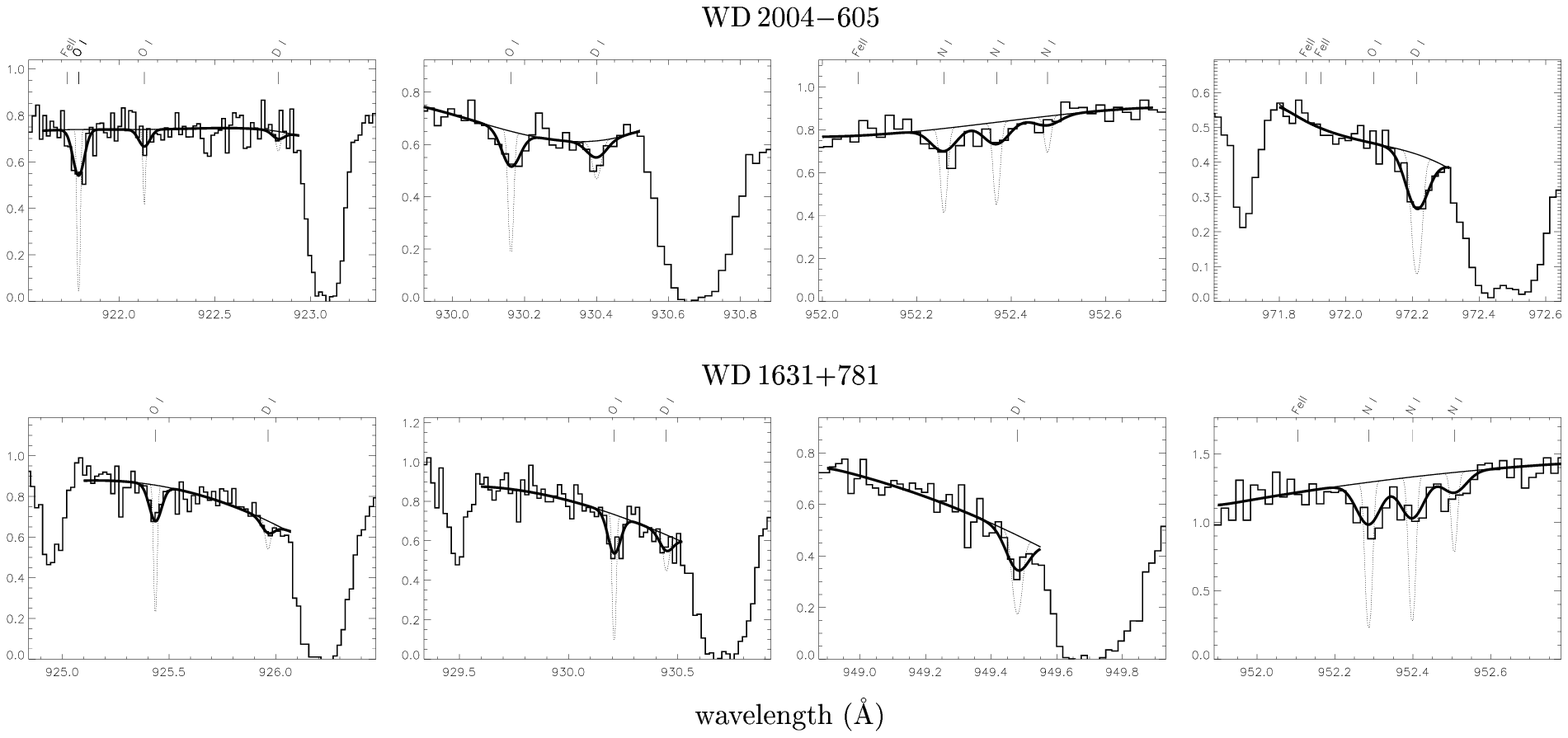,height=11cm}
\caption{Examples of \fuse\ spectral windows fitted on the lines 
of sight of the DA1 white dwarfs WD$\,$2004$-$605 and WD$\,$1631$+$781. 
The list of all the \ion{D}{1}, \ion{O}{1}, and \ion{N}{1} 
lines included in the fits are reported in Table~\ref{table_lines}. 
Histogram lines are the data, the solid lines are the 
fits and continua, and the dotted lines are the model profiles prior to 
convolution with the LSF. 
Y-axis is flux in $10^{-12}\,$erg/cm$^2$/s/\AA.
\label{wd2004_wd1631_fit_plot}}
\end{figure}

\begin{figure}
\psfig{file=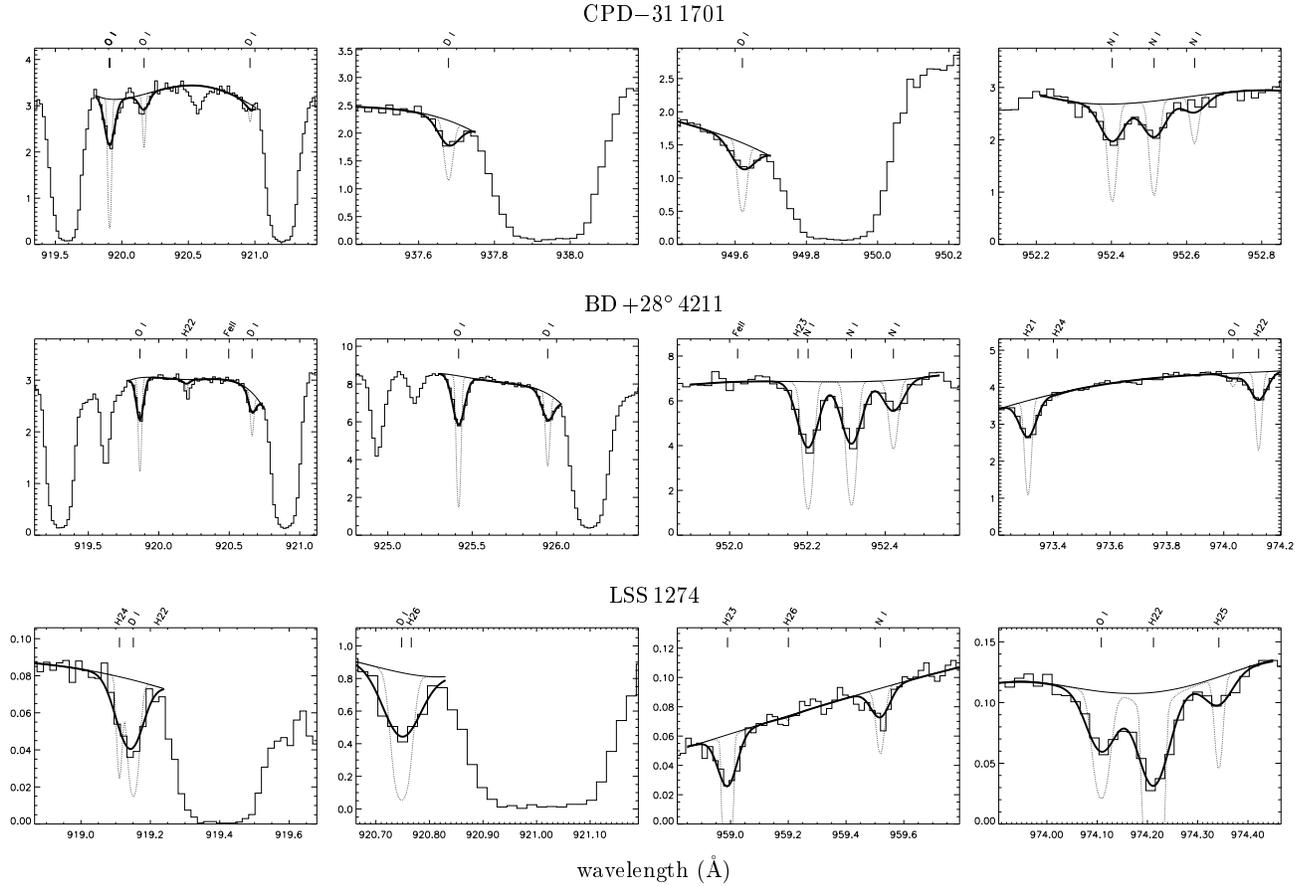,height=14cm}
\caption{Examples of \fuse\ spectral windows fitted on the lines 
of sight of the sdO subdwarfs 
CPD$-$31$\,$1701, BD$\,+28^{\circ}\,4211$, and LSS$\,$1274. 
The list of all the \ion{D}{1}, \ion{O}{1}, and \ion{N}{1} 
lines included in the fits are reported in Table~\ref{table_lines}. 
Histogram lines are the data, the solid lines are the 
fits and continua, and the dotted lines are the model profiles prior 
to convolution with the LSF. 
The H$_2$ lines of the levels $J=2$ to $J=6$ are noted H22 to H26. 
Y-axis is flux in $10^{-11}\,$erg/cm$^2$/s/\AA.
\label{cpd_bd_lss_fit_plot}}
\end{figure}

Spectral windows (see examples in
Figures~\ref{sirius_fit_plot}-\ref{cpd_bd_lss_fit_plot}) including
\ion{D}{1}, \ion{O}{1}, \ion{N}{1}, and other species 
transitions were extracted 
from the \fuse\ spectra available for each of the different datasets 
(Table~\ref{table_obslog}) and fitted simultaneously
with the \ow\ procedure. This software, which has been
developed by Martin Lemoine and the French \fuse\ Team, finds the best
Voigt profiles compatible with all the spectral windows, using \kid\ 
minimization. Several parameters are free to vary during the fitting
procedure, including the column densities, the radial velocities of
the interstellar clouds, their temperatures and turbulent velocities,
and the shapes of the stellar continua, which are modeled by low
order polynomials. 

Some instrumental parameters are also free to vary, including the
spectral shifts between the different spectral windows (in order to
correct for inaccuracies of the \fuse\ wavelength calibration) and the
widths of the Gaussian Line Spread Functions (LSF) used to convolve
with the Voigt profiles (the \fuse\ LSF is not determined accurately,
but it probably depends on the channel and the wavelength).  The fits
include only unsaturated lines in order to reduce uncertainties
resulting from saturation effects and other systematic errors. 
The plots of the fits prior to convolution with the LSF give us the
saturation criterion, following the method used by H\'ebrard et
al.~(\citealp{hebrard02}): all the lines with un-convolved absorption
profiles reaching close to the zero flux level were not included in
the final fits.
As \ion{D}{1}, \ion{O}{1}, and \ion{N}{1} transitions are numerous in
the \fuse\ bandpass, with a large range of oscillator strengths, it 
was possible to fit numerous unsaturated lines for most of the targets
(see Table~\ref{table_lines}).

The laboratory wavelengths and
oscillator strengths are from Morton~(\citealp{morton91,morton02}). 
When needed, those of H$_2$ were added from a compilation by E. Roueff 
(Abgrall et al.~\citealp{abgrall93a}, \citealp{abgrall93b}). 
For the lines that we used, no biases due to oscillator strength 
uncertainties were found (see tests on oscillator strength values 
in H\'ebrard et al.~\citealp{hebrard02}).
Only one interstellar component was assumed for each target. 
This assumption is unlikely to be true for many lines of sight, but 
it will have no effect on the column densities measured from unsaturated 
lines (see tests of this assumption in H\'ebrard et 
al.~\citealp{hebrard02}).
Thus, we report total column densities, integrated along each line 
of sight.  

The simultaneous fitting of numerous unsaturated lines allow
statistical and systematic errors to be reduced, especially those due
to continuum placements, LSF uncertainties, line blending, flux and
wavelength calibrations, and atomic data uncertainties. The 1-$\sigma$
error bars were computed using the \deltakid\ method; both statistical
and systematic effects were taken into account. The details of the
\deltakid\ method may be found in H\'ebrard et
al.~(\citealp{hebrard02}). 

The \ion{D}{1}, \ion{O}{1}, and \ion{N}{1} lines included in the fits
are reported in Table~\ref{table_lines}.  Note that due to the
redundancy of the \fuse\ wavelength coverage (and to the use of
several slits for some targets), a given transition may be observed
several times. These independent observations are all included in the
fits, allowing the reduction of the error bars, and identification of
possible instrumental artifacts.

\section{Individual targets}
\label{Individual_targets}

We discuss here the seven individual targets with new \fuse\ results.

\subsection{Sirius~B}

\ion{D}{1} and \ion{N}{1} interstellar column densities were measured 
from HST-GHRS spectra by H\'ebrard et al.~(\citealp{hebrard99}) for
the two interstellar components detected along the line of sight to
the white dwarf Sirius~B. Using these results, we adopt the following
values for the total column densities integrated along the line on
sight (\ie\ for the sum of the two components), with \1s\ error bars:
$\log N$(\ion{D}{1})$\,=12.88\pm0.08$ and 
$\log N$(\ion{N}{1})$\,=13.35\pm0.03$ (\cmmd).

The \ion{O}{1} column density reported by 
H\'ebrard et al.~(\citealp{hebrard99}) was obtained using only
the saturated 1302~\AA\ line and was inaccurate. 
In addition, the fit of this line was poor. The new \ion{O}{1} 
column density reported here was obtained from a 
public set of \fuse\ GI observations 
of Sirius~B (Holberg et al.~\citealp{holberg02}), which show unsaturated 
\ion{O}{1} absorption lines (see Table~\ref{table_lines}).
Two examples of fits of the \ion{O}{1} lines are given in 
Figure~\ref{sirius_fit_plot}.

\subsection{WD$\,$2004$-$605}

The photometric distance of the DA1 white dwarf WD$\,$2004$-$605
reported by Holberg et al.~(\citealp{holberg98}) is 52~pc. Three
observations of this target were performed as part of the \fuse\ 
Science Team D/H program.  Some exposures were lost due to thermal
misalignment of the instrument (for example, the ten first exposures
of P2042202), but the final combined spectra have a good signal to 
noise ratio, with a stellar continuum around
$1-4\times10^{-12}\;$erg$\,$cm$^{-2}\,$s$^{-1}\,$\AA$^{-1}$.  Numerous
unsaturated \ion{D}{1}, \ion{O}{1}, and \ion{N}{1} lines were
detected. Neither lines from stellar \ion{He}{2} nor interstellar
H$_2$ were detected; thus, the fitting process is quite simple.
Examples of four spectral windows are shown in
Figure~\ref{wd2004_wd1631_fit_plot}-upper panel.
\ion{O}{1} and \ion{N}{1} column densities were measured independently 
using the curve-of-growth method (Lehner et
al.~\citealp{lehner03}).  Both methods (profile fitting and 
curve-of-growth) gave similar results, and the final values
reflect the combined effort of these two~analyses.

\subsection{WD$\,$1631$+$781}

The photometric distance of WD$\,$1631$+$781 is 67~pc (Holberg et
al.~\citealp{holberg98}).  Observed in January 2000, this DA1 white
dwarf was the first \fuse\ target for which well resolved \lyb\
\ion{D}{1} lines were detected.  However, \lyb\ was the only
\ion{D}{1} transition observed because, at that time, uncontrolled
thermal misalignment of the instrument prevented acquisition of the
targets in the MDRS slits of the SiC channels. A new observation
performed one year later obtained the shorter wavelength Lyman lines
in the SiC channels. Combined spectra of the two observations show
numerous unsaturated lines, absorbed on a
$\sim1-5\times10^{-12}\;$erg$\,$cm$^{-2}\,$s$^{-1}\,$\AA$^{-1}$ simple
stellar continuum. No strong H$_2$ lines were detected, so line
blending is not an issue. Four examples of fits are plotted in
Figure~\ref{wd2004_wd1631_fit_plot}-lower panel. Here again, the
\ion{O}{1} and \ion{N}{1} column densities include input from 
the independent work by Lehner et al.~(\citealp{lehner03}).

\subsection{CPD$-$31$\,$1701}

Subdwarfs can yield a high UV flux at distances beyond 100~pc. Thus,
they allow the diffuse interstellar medium to be observed on larger
path lengths and higher column densities compared to white dwarf sight
lines. CPD$-$31$\,$1701 is a sdO subdwarf which is $131\pm28$~pc from
the Sun, according the Hipparcos parallax measurement. This target was
observed as part of the \fuse\ Science Team D/H program, through both
the LWRS and the MDRS slits. The stellar flux is relatively high,
ranging from $2$ to
$5\times10^{-11}\;$erg$\,$cm$^{-2}\,$s$^{-1}\,$\AA$^{-1}$.  
Numerous unsaturated \ion{D}{1}, \ion{O}{1}, and \ion{N}{1} lines were
detected in absorption.  Some H$_2$ lines were also detected, but are
excluded from the analysis, due to blends with atomic
lines. Similarly, we used only interstellar lines that were not
strongly blended with the numerous stellar lines in the spectra.

Comparisons made with stellar models show that blending of the
interstellar lines used with stellar lines are not a significant
problem, and that fitted polynomials produce acceptable estimates of
the continua. An extensive study of the possible effects of stellar
models on interstellar column density measurements for lines of sight
toward subdwarfs will be the aim of a forthcoming paper. The error
bars reported here should be considered as conservative first
estimates, and will be refined in a future report. Four examples of
fits are given in Figure~\ref{cpd_bd_lss_fit_plot}-upper panel.

\subsection{BD$\,$+28$^{\circ}$4211}

BD$\,+28^{\circ}\,4211$ is a bright sdO subdwarf, located $\sim~100$~pc 
from the Sun, \ie\ at the limit of the Local Bubble (Snowden et
al.~\citealp{snowden98}; Ferlet~\citealp{ferlet99}; Sfeir et
al.~\citealp{sfeir99}).  The first analysis of \fuse\ spectra of this
target was published by Sonneborn et al.~(\citealp{sonneborn02}), who
reported high D/O and low O/H ratios compared with the other \fuse\ 
lines of sight (Moos et al.~\citealp{moos02}). The O/H was also
significantly lower than the ratio found by Meyer et
al.~(\citealp{meyer98}) for the local interstellar medium.  These
differences may result from local inhomogeneities in the abundance of
oxygen toward that target, but also from under-evaluation of the
\ion{O}{1} column density (or both). The error bar on
$N($\ion{O}{1}$)$ reported by Sonneborn et al.~(\citealp{sonneborn02})
was large even though the signal-to-noise ratio was high, primarily
because the three \ion{O}{1} transitions used are near saturation. As
can be seen in their Figure~11, the strongest line
($\lambda\,930.3\,$\AA) is the best fitted, while the fit is poor for
the less saturated line ($\lambda\,919.9\,$\AA). We re-investigated
this analysis in order to reduce the error bars, especially on
$N($\ion{O}{1}$)$.

Six extra observations of BD$\,+28^{\circ}\,4211$ were performed using
all three \fuse\ slit widths (LWRS, MDRS, and HIRS) in July 2001.
This added an extra $16.5\times10^3$~s of exposure time to the
$51.6\times10^3$~s of exposure time for the four observations used by
Sonneborn et al.~(\citealp{sonneborn02}).  Although the increase in
exposure time is modest, it includes new spectra obtained through the
narrowest slit (HIRS).  All of the data were processed (or
reprocessed) using CalFUSE~2.1. Spectral lines obtained through the
three slits and the four channels were fitted simultaneously; the
final fit includes 88 spectral windows 
(four of them are plotted in Figure~\ref{cpd_bd_lss_fit_plot}-middle 
panel.)

As some \ion{D}{1} and \ion{O}{1} absorption lines are blended with
H$_2$ lines (levels $J=1$, $2$, and $3$), a molecular component was
added in the fit, with all of the molecular parameters remaining free
to vary.  Thus, the final error bars on \ion{D}{1} and \ion{O}{1}
column densities take into account the uncertainties in the H$_2$
column densities.  The \ion{D}{1} and \ion{N}{1} lines included in our
new fits were essentially the same as the ones used in the initial
analysis -- some extra lines were included but they are weak. The main
difference with the analysis of Sonneborn et
al.~(\citealp{sonneborn02}) concerns the \ion{O}{1} lines. The
transition $\lambda\,930.3\,$\AA, which is probably slightly
saturated, was not included and we added two extra unsaturated lines,
$\lambda\,922.2\,$\AA\ and $\lambda\,974.1\,$\AA.

Our new results agree with the initial ones by Sonneborn et
al.~(\citealp{sonneborn02}) at a \1s\ level. The error bars are
reduced however, and this sight line now appears to be more in
agreement with the other ones in terms of D/O and O/H.

\begin{figure}
\psfig{file=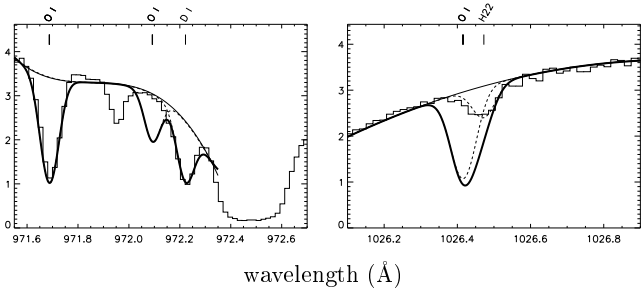,height=4.5cm}
\caption{Inaccurate tabulated \ion{O}{1} $f$-values. The \ion{O}{1} 
lines 
at 972.14~\AA\ and 1026.47~\AA\ (wavelengths at rest) observed toward 
BD$\,+28^{\circ}\,4211$ are shown here, according to the solution 
found with the other lines and the $f$-values tabulated by 
Morton~(\citealp{morton91}). Histogram lines are the data, the solid 
lines are the computed profiles and continua, and the dashed lines 
are the computed profiles for each species (all convolved with 
the LSF). Y-axis is flux in $10^{-11}\,$erg/cm$^2$/s/\AA.
The clear detection of other lines 
(\ion{O}{1}, \ion{D}{1}, and H$_2$ $J=2$) on the 
two spectral windows allows wavelength shifts to be determined. 
The line at $\lambda\,971.95\,$\AA\ is unidentified. 
The two tabulated $f$-values are too large at 
least by factors of 5 and 30, respectively. These two lines were 
not used toward any target in the present study.
\label{fig_bad_OI_F-values}}
\end{figure}

Since the BD$\,+28^{\circ}\,4211$ line of sight presents both a high
column density and high signal-to-noise ratio observations, it allows
inaccurate tabulated $f$-values of \ion{O}{1} lines to be
identified. The two \ion{O}{1} triplets at 972.14~\AA\ and
1026.47~\AA\ are plotted in Figure~\ref{fig_bad_OI_F-values}.  Fits
were performed following the prescription discussed above,
\ie\ without using these two lines. We see 
that the $f$-values tabulated by Morton~(\citealp{morton91}) for the
two lines are clearly overestimated, at least by a factor of 5 and 30
for the $\lambda\,972.14\,$\AA\ and $\lambda\,1026.47\,$\AA\ lines,
respectively.  The $f$-values for these two weak intersystem
transitions were calculated by Kurucz \&
Peytremann~(\citealp{kurucz75}) and show large error bars. We did not
use them for column density measurements toward any target.

\subsection{LSS$\,$1274}

LSS$\,$1274 is a sdO subdwarf, located $580\pm100$~pc from the Sun
(Dreizler~\citealp{dreizler93}). Thus, with HD$\,$191877 and
HD$\,$195965 (Hoopes et al.~\citealp{hoopes03}), this is one of the
three longest lines of sight for which deuterium abundances have been
measured with \fuse. This star is fainter than the three other
subdwarfs presented here; the stellar flux ranges from $1$ to
$2.5\times10^{-12}\;$erg$\,$cm$^{-2}\,$s$^{-1}\,$\AA$^{-1}$.  Numerous
interstellar lines are detected but most of them are saturated due to
the high column density. In addition, numerous H$_2$ interstellar
lines and stellar features are present.  We used only the interstellar
lines un-blended with stellar features. We included the following
species in the fit: \ion{D}{1}, \ion{O}{1}, \ion{N}{1}, \ion{Fe}{2}, 
and the levels $J=1$ to $J=5$ of H$_2$. Four examples of fits are 
plotted in Figure~\ref{cpd_bd_lss_fit_plot}-lower panel.

Finally, $\lambda\,974\,$\AA\ is the only unsaturated \ion{O}{1} line
that we can use. This makes our \ion{O}{1} column density especially
dependent on the $f$-value uncertainty. This is probably the largest
value of $N($\ion{O}{1}$)$ measurable with \fuse.  Note that
\ion{O}{1} lines with lower $f$-values within the \fuse\ range are 
located at wavelengths shorter than 925~\AA, where they are usually
undetectable due to blends with the strong \ion{H}{1} Lyman lines.
Thus, $\log N$(\ion{O}{1}) values larger than 17.9 probably will not be
measurable with \fuse, and will require the $\lambda\,1356\,$\AA\ line
in the HST wavelength range.

\subsection{Feige$\,$110}

\begin{figure}
\psfig{file=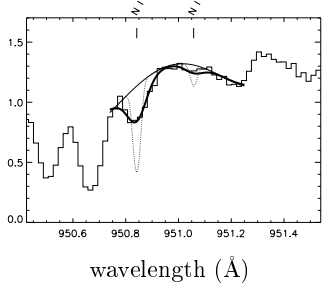,height=6.2cm}
\caption{\ion{N}{1} fit toward Feige$\,$110. 
Histogram lines are the data, the solid lines are the 
fits and continua, and the dotted lines are the model profiles prior to 
convolution with the LSF. 
Y-axis is flux in $10^{-11}\,$erg/cm$^2$/s/\AA.
\label{feige_fit_plot}}
\end{figure}

An extensive study of the Feige$\,$110 sight line was performed by
Friedman et al.~(\citealp{friedman02}). We use their \ion{D}{1} and
\ion{O}{1} column densities, together with a new $N$(\ion{N}{1}) 
value. This latter one was obtained from the data set previously used
by Friedman et al.~(\citealp{friedman02}).  Due to the numerous
stellar and H$_2$ interstellar lines, most of the \ion{N}{1} lines are
blended so they do not allow a column density measurement.  However,
two unsaturated \ion{N}{1} lines are detected around 951~\AA, which
seem un-blended. The corresponding fit is plotted in
Figure~\ref{feige_fit_plot}. Only the strongest one is clearly
detected. Our result is therefore subject to possible systematic
errors, due to an uncontrolled blend or a poorly known oscillator
strength. Note that we did not see any of these effects in the case of
BD$\,+28^{\circ}\,4211$, for which these lines were used, together
with other \ion{N}{1} lines.

\section{Results}
\label{Results}

The measured column densities as well as the D/O and D/N ratios are
displayed in Table~\ref{table_results}, sorted from the lowest
\ion{D}{1} column density to the highest.  In addition to the seven
targets discussed above, the Table includes 12 targets with previously
published \fuse\ analyses (HZ$\,$43$\,$A, G191$-$B2B, Capella,
WD$\,$0621$-$376, WD$\,$2211$-$495, WD$\,$1634$-$573,
WD$\,$2331$-$475, GD$\,$246, HZ$\,$21, 
Lan$\,$23, HD$\,$191877, and HD$\,$195965), and five other targets
with {\it IMAPS}, HST, and/or {\it Copernicus} observations
($\alpha\,$Vir, 
$\delta\,$Ori$\,$A, $\gamma\,$Cas, $\epsilon\,$Ori, and
$\iota\,$Ori).  All the references are reported in
Table~\ref{table_results}.  Note that some of the column densities for
the four latter targets were obtained using only one transition, which
make them more dependent on possible inaccuracies in the atomic
data. For the case of the HST measurements of $N$(\ion{O}{1}) reported
by Meyer et al.~(\citealp{meyer98}), we corrected the values with an
updated oscillator strength for the transition used
($\lambda\,1356\,$\AA), in agreement with
Meyer~(\citealp{meyer01}). Altogether, a sample of 24 lines of sight
with \ion{D}{1}, \ion{O}{1}, and \ion{N}{1} interstellar column
densities is available. We did not include lines of sight with
published D/H but without known D/O.

York~(\citealp{york83}) has measured the column density for \ion{D}{1}, 
\ion{O}{1}, and \ion{N}{1} towards $\lambda\,$Sco.  The total D/O value 
is low but presents large variations from component to component.
There is a possibility the uncertainties should be larger due to the
use of saturated lines.  In addition, although the star is
$\sim220$~pc (Hipparcos parallax measurement), the Local Bubble wall
has protrusions and appears porous in that general direction (Lallement
et al.~\citealp{lallement03}),
so the location of the gas measured is quite uncertain.
Therefore, we have not included this line of
sight in our sample.

\begin{figure}
\psfig{file=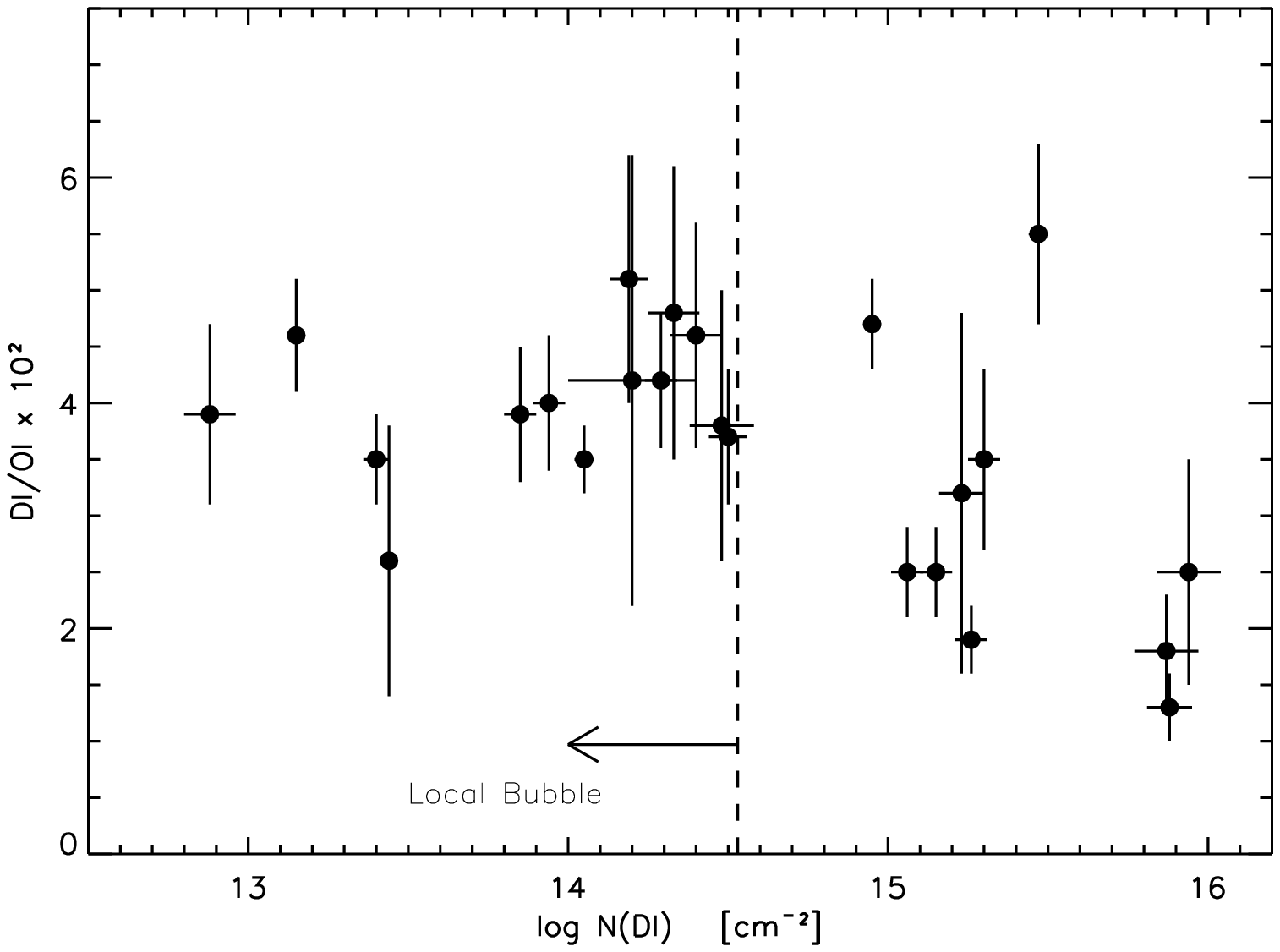,height=6.5cm}
\psfig{file=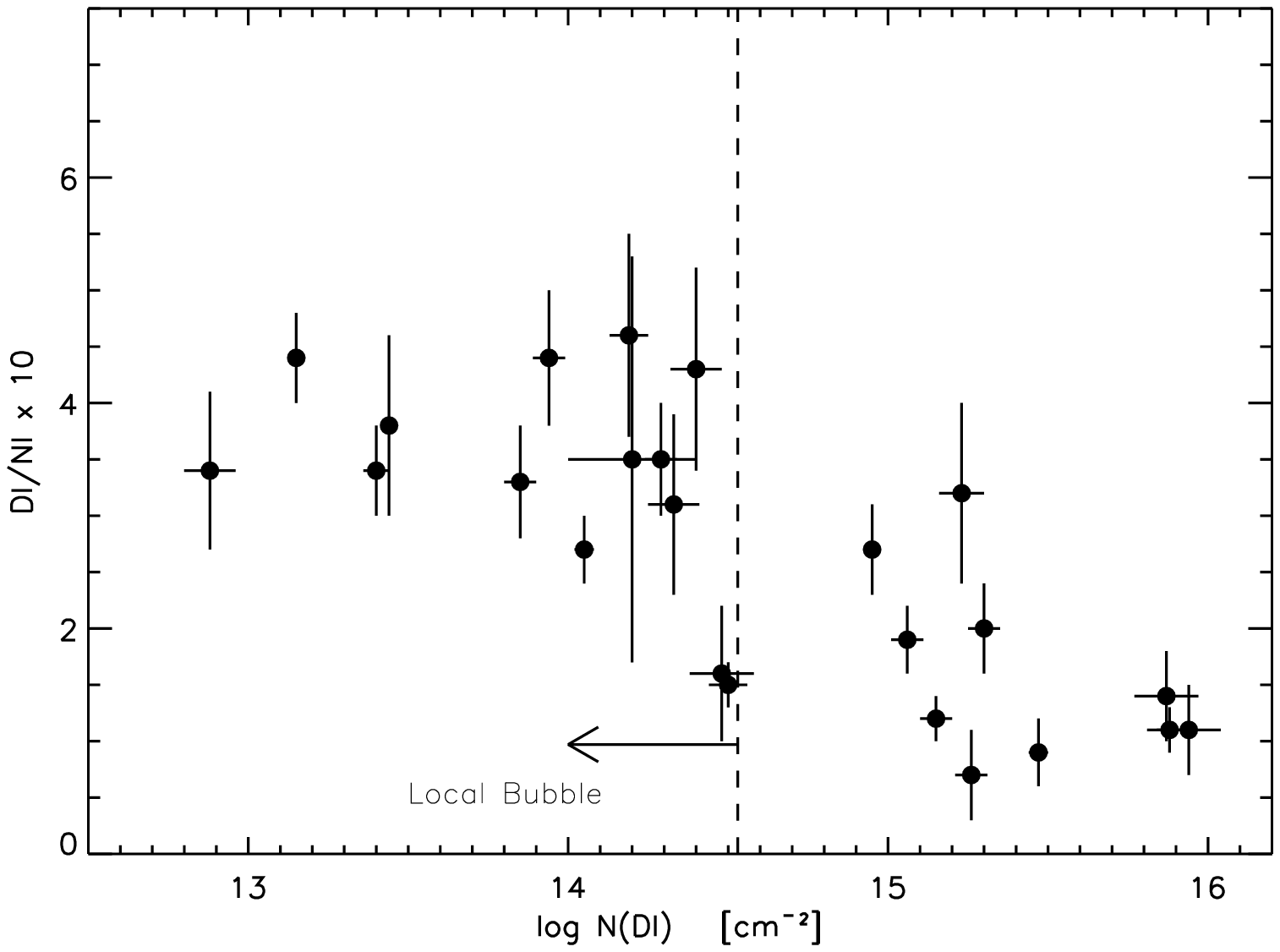,height=6.5cm}
\caption{D/O and D/N measurements in the interstellar medium as a 
function of the \ion{D}{1} column density. 
The plotted error bars~are~$1\,\sigma$. 
The dotted line indicates the limit of the Local Bubble, 
inside which the D/O ratio is homogeneous.
\label{fig_dso_and_dsn}}
\end{figure}

Figure~\ref{fig_dso_and_dsn} presents the D/O and D/N ratios as a
function of the \ion{D}{1} column density. For the targets with the
lowest \ion{D}{1} column densities, \ie\ for the near local
interstellar medium (LISM), the D/O ratio appears to be remarkably
constant, contrary to D/N. For the 14 targets with $\log
N($\ion{D}{1}$) \leq 14.5$, we obtained the weighted mean
D/O$\,=(3.84\pm0.16) \times 10^{-2}$ 
(all the means reported here are weighted by the reported errors).
The $\pm0.16$ error that we
report here and the errors that we report below are $1\,\sigma$
uncertainties in the mean. Note that the square root of the weighted
average variance, which may be used to compare a single new
measurement with the mean, is $\pm0.49$.  The
\kid\ for this weighted mean is 8.4 for 13 degrees of freedom; the  
data are consistent with a single value for the the D/O ratio in
the near LISM, which is probably the Local Bubble.  In comparison, for
the same 14 targets, the \kid\ for the weighted mean of D/N is 37.3,
indicating that D/N varies more
significantly from one line of sight to the
other. At higher \ion{D}{1} column densities, both the D/O and D/N
ratios vary.  
The weighted means and the \kid\ for all the sight lines are 
respectively $3.05\times 10^{-2}$ and 117.9 for D/O, and 
$1.90\times 10^{-1}$ and 189.9 for D/N, for 23 degrees of freedom.

\begin{figure}
\psfig{file=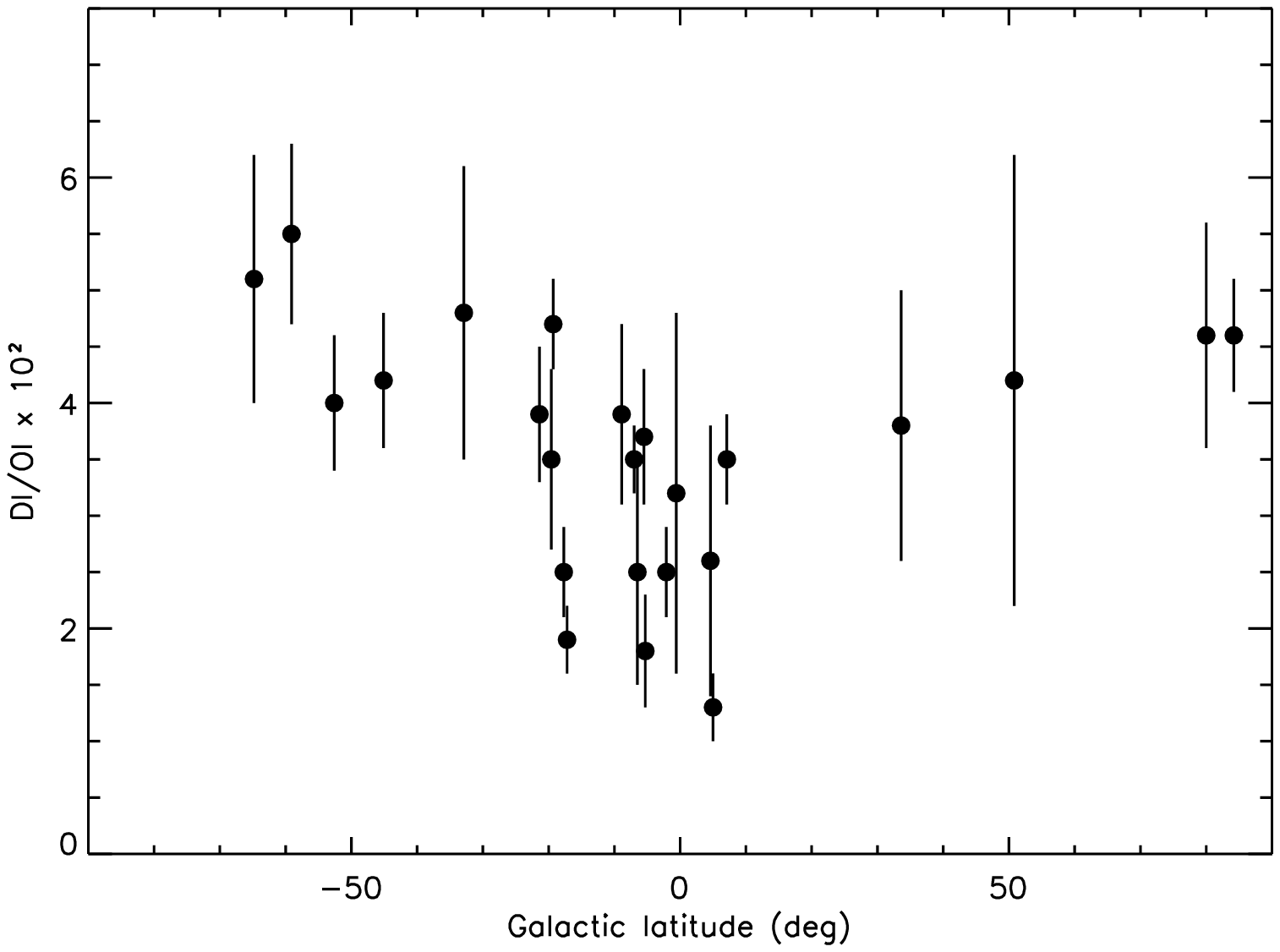,height=6.5cm}
\psfig{file=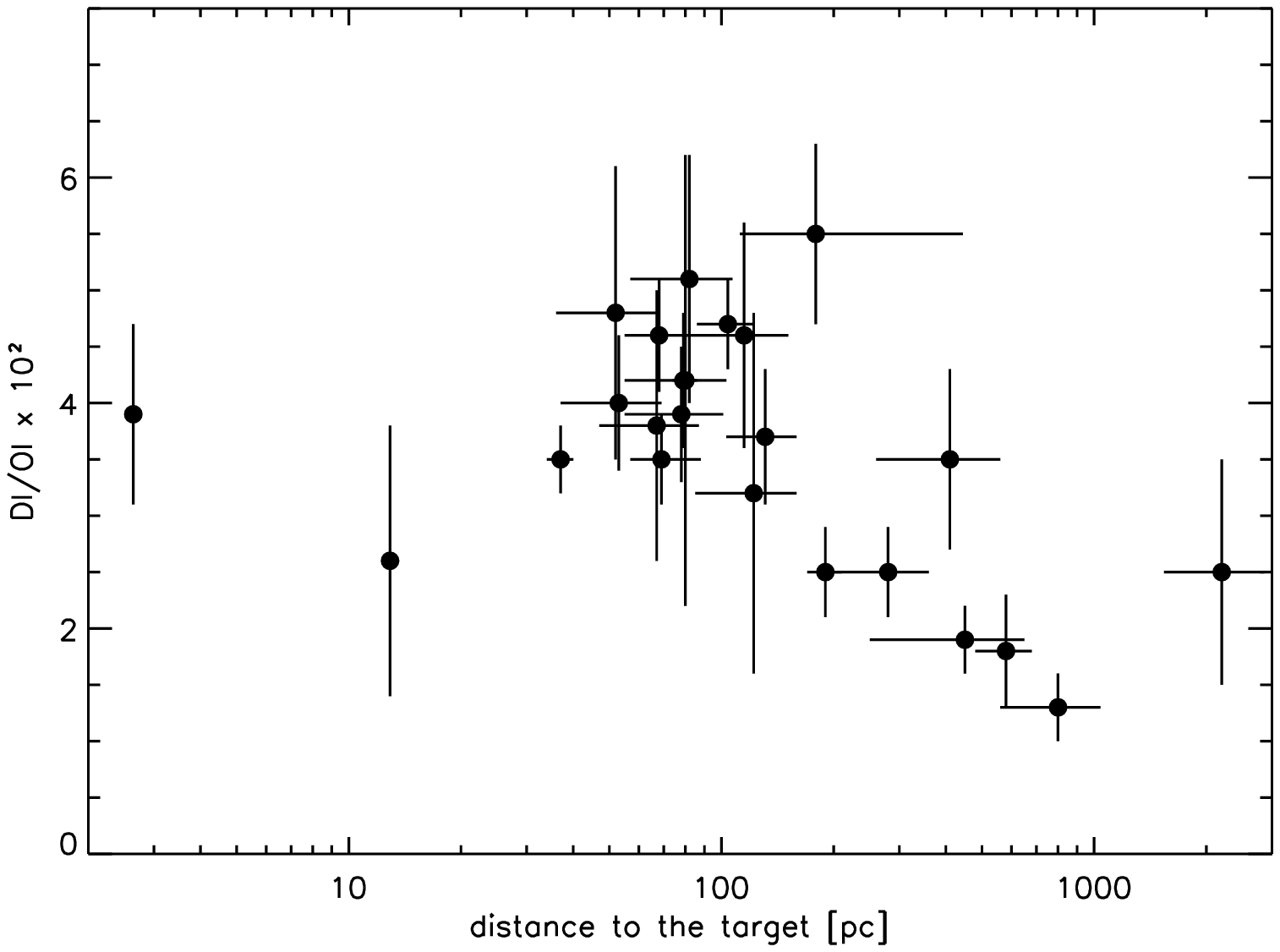,height=6.5cm}
\caption{D/O measurements in the interstellar medium 
as a function of the Galactic latitude $b$ and the distance 
of the targets.  
The plotted error bars are~$1\,\sigma$.
\label{fig_bII_and_dist_dso}}
\end{figure}

In order to better examine variations of the D/O ratio in different
regions of the LISM, we plot it as a function of the Galactic latitude
and distance in Figure~\ref{fig_bII_and_dist_dso}.  The sample
contains only a few high latitude stars and most of them are close
with modest \ion{D}{1} column densities.  Thus, it is not surprising
that the scatter is limited.  However, the D/O values near the
Galactic plane, which correspond to a wide range of \ion{D}{1} column
densities, show a larger scatter, in agreement with the variation as a
function of column density discussed above.  As a function of the
distance, D/O is homogeneous on lines of sight toward stars of 150~pc
or less.  For the 16 targets in that range, we find 
D/O$\,= (3.96\pm0.15) \times 10^{-2}$, with a \kid\ of 12.5 for 15
degrees of freedom. So, again, there is no evidence for variation of
D/O in the LISM, even to distances slightly outside the Local Bubble.

\begin{figure}
\psfig{file=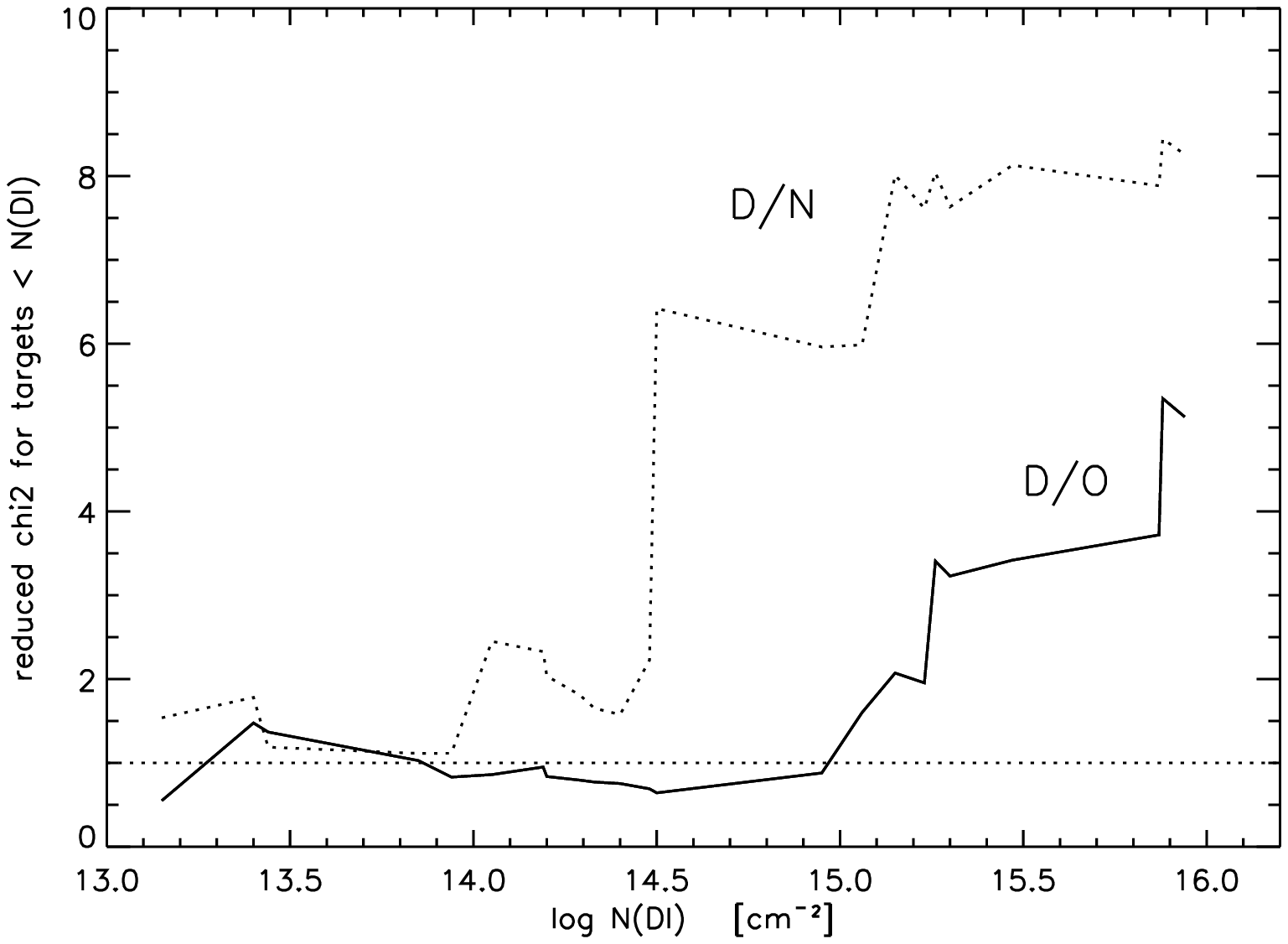,height=6.5cm}
\psfig{file=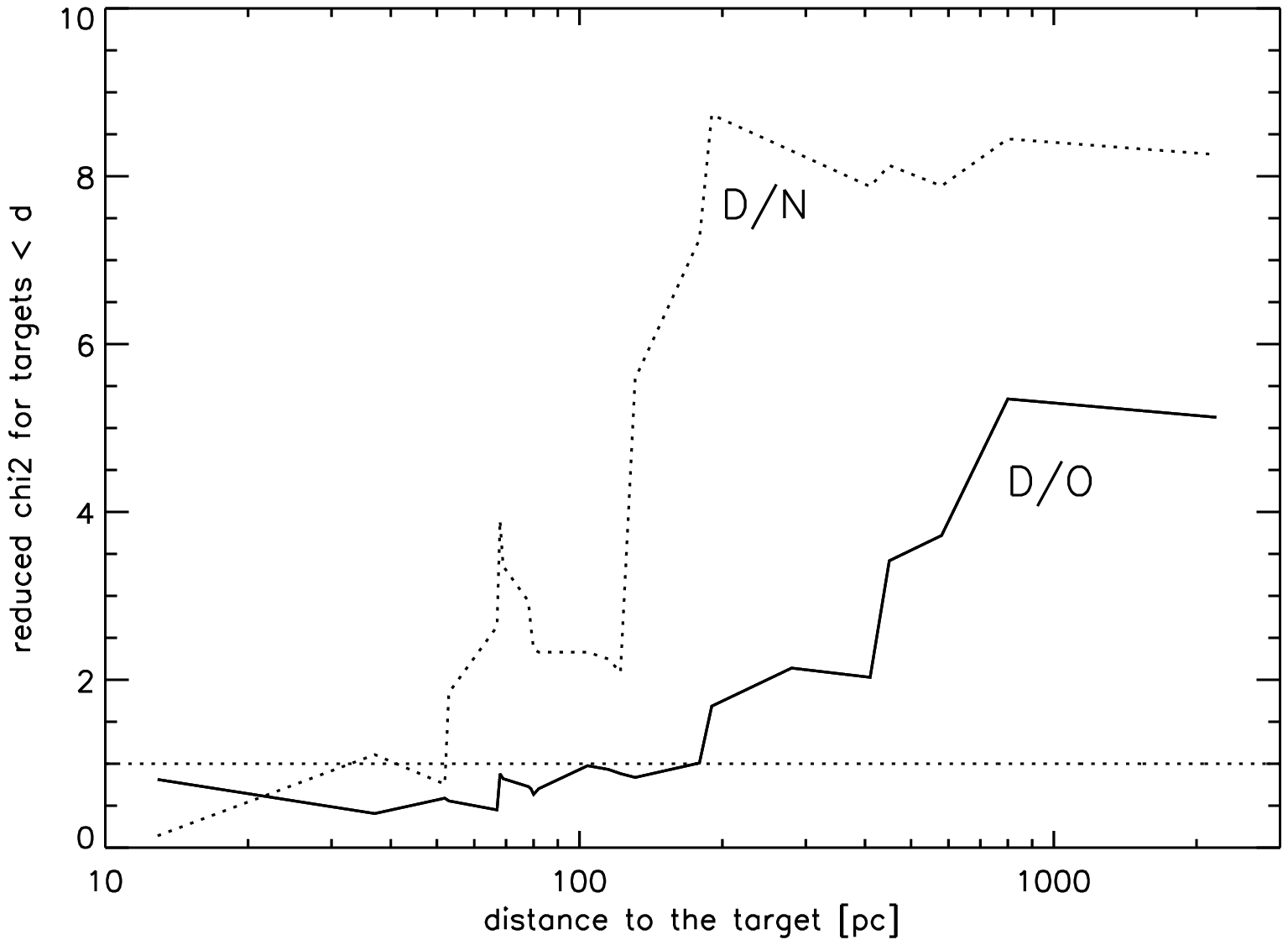,height=6.5cm}
\caption{The reduced \kid\ of the D/O and D/N weighted mean as a 
function of the \ion{D}{1} column density and distance upper 
limits assumed to 
define the sample used for the weighted mean calculation (see text).
D/O is homogeneous for $d<150$~pc and $\log N($\ion{D}{1}$)<14.5$; 
it shows significant spatial variations beyond these limits. 
D/N varies even for low $\log N($\ion{D}{1}$)$ and $d$ values. 
\label{fig_chi}}
\end{figure}

We plot in Figure~\ref{fig_chi} the reduced \kid\ of the D/O weighted
mean as a function of the \ion{D}{1} column density and as a function
of the target distances, which serve as upper limits for the distance
to the ISM gas. The \kid\ values remain similar to the degrees of
freedom (reduced \kid\ near unity) for distances lower than 150~pc and
$N($\ion{D}{1}$)<15$, but the values increase significantly above
these limits as the D/O values deviate more and more from the 
low-distance and low-column density mean. We can not draw strong
conclusions about D/O variations or homogeneity in the range
$14.5<\log N($\ion{D}{1}$)<15$, because we have only one target
(BD$\,+28^{\circ}\,4211$). Thus, we adopt $\log N($\ion{D}{1}$) \leq
14.5$ as a conservative cutoff for the D/O homogeneity. Using a D/H
ratio ranging between $1.3$ and $1.5\times10^{-5}$ (see
Sect.\ref{dsh_value_inferred_from_dso}), this cutoff corresponds to
$\log N($\ion{H}{1}$)=19.3-19.4$, which is the value used by Sfeir et
al.~(\citealp{sfeir99}) for the wall of the Local Bubble. In 
comparison to D/O, D/N shows variability at lower distances and
\ion{D}{1} column densities (Figure~\ref{fig_chi}). The D/N variations
are actually detected as soon as a few targets are included in the
sample.  We have also examined the reduced \kid\ starting with the
maximum values of \ion{D}{1} column density and target distances,
\ie\ from right to left.  For both D/O and D/N, after a few points the
reduced \kid\ takes on a value greater than one, indicating variability
at large distances and column densities.

Some of the distances are photometric, with no error bars reported and
should be regarded as uncertain. In Figure~\ref{fig_bII_and_dist_dso},
we assumed a 30~\% 
uncertainty for such cases, but this may be an underestimation. 
For example, in the case of CPD$-$31$\,$1701, Bauer
\& Husfeld~(\citealp{bauer95}) reported a photometric distance of
$320\pm80$~pc, while the Hipparcos measured parallax of
$7.62\pm1.51$~mas is equivalent to $131\pm28$~pc. Moreover, the
distances are those of the stars, and are only upper limits for the
distances to the interstellar gas. We prefer to define the local
interstellar medium in terms of column density. Thus, we adopt the D/O
ratio obtained with the \ion{D}{1} column density criterion rather
than distance criterion, and we report the following value for the D/O
ratio in the Local Bubble from \fuse\ observations with 
$\log N($\ion{D}{1}$) \leq 14.5$:

$$({\rm D/O})_{\rm{LB}}= (3.84\pm0.16) \times 10^{-2}  \ \ 
(1\,\sigma).$$

This result, which is obtained from 14 lines of sight, is in agreement
with the earlier \fuse\ results: Moos et al.~(\citealp{moos02}) and
Oliveira et al.~(\citealp{oliveira03}) reported 
$({\rm D/O})_{\rm{LB}}=(3.76\pm0.20)\times 10^{-2}$ and 
$({\rm D/O})_{\rm{LB}}=(3.87\pm0.18)\times
10^{-2}$ for 5 and 8 lines of sight, respectively. The increased 
number of sight lines further reduces the uncertainty.

\section{Discussion}
\label{Discussion}

We discuss here the above results. The various deuterium abundance 
values discussed below are summarized in Table~\ref{table_sum}. 

\subsection{D/O and D/H homogeneities in the LISM}

The \ion{D}{1} and \ion{O}{1} column densities differ by less than two
orders of magnitude. Hence, as discussed previously, systematic errors
in D/O 
are likely to be smaller than those associated with the determination
of D/H. Moreover, any residual systematic over- or underestimation of
our column densities, due for example to subtle effects in the LSF,
zero flux level estimates, or line of sight velocity structures, will
be reduced when taking the ratio of $N$(\ion{D}{1}) to
$N$(\ion{O}{1}).  According to the above effects, this is 
probable that part of the errors
for $N$(\ion{D}{1}) and $N$(\ion{O}{1}) are correlated.  However 
the errors are combined with the conservative assumption they are
independent. This tends to slightly overestimate the error bars on D/O
and may be why the reduced \kid\ for the Local Bubble is less than
one. Thus, it is likely that the results for D/O are robust.

The homogeneity of D/O in the near LISM argues against variations of
D/H in the Local Bubble.  Indeed, the only possible way that D/O could
be stable while D/H varies, would be for D/H and O/H to be
correlated. Moreover, they would have to vary in a precise manner that
allowed D/O to remain constant.  That seems improbable for two 
{\it different} reasons: (i) O/H appears to be uniform in the ISM over
paths of several hundred parsecs (\eg,~Meyer et al.~\citealp{meyer98}), 
and (ii) astration processes should lead to an anti-correlation of
\ion{D}{1} and \ion{O}{1} abundances. Thus, the stability of D/O in 
the near LISM is a strong argument for the stability of both D/H 
and O/H in the Local Bubble.

Conversely, the variations of D/O from one sight line to the other for
distances greater than 150~pc and \ion{D}{1} column densities larger
than $10^{15}$~\cmmd\ is an argument which supports D/H variations at
these scales, as studies by Meyer et al.~(\citealp{meyer98}),
Cartledge et al.~(\citealp{cartledge01}), or Andr\'e et
al.~(\citealp{andre03}) show that the O/H ratio does not vary
significantly over the distances covered in this study.  Thus, while
the interstellar medium seems well mixed inside the Local Bubble,
this is not the case beyond 150~pc.  
It appears that D/O varies from sight line to sight line,
whereas the studies above show that O/H does not. A comparison
between the interstellar values and the low-metallicity quasar sight
lines values shows that \ion{D}{1} and \ion{O}{1} are, at least
partially, anti-correlated over large ranges in the metallicity.
 However, for most of the distant ISM 
targets studied here, the mechanisms affecting \ion{D}{1} do not 
appear to affect \ion{O}{1} significantly. 
This will provide a constraint on mixing models (\eg, de Avillez \& 
Mac Low~\citealp{deavillez02}) and chemical evolution models
(\eg, Chiappini et al.~\citealp{chiappini02}).

Because of the variations in the D/O ratio outside of the Local
Bubble, it is difficult to choose which value for D/O is
representative of the present epoch.  There certainly is no reason to
preferentially adopt the D/O value measured in the Local Bubble. Note
a possible trend; the D/O values are lower for the higher \ion{D}{1}
column densities and especially the larger distances
(Figures~\ref{fig_dso_and_dsn} and \ref{fig_bII_and_dist_dso}). 
Consider the three farthest targets, both in terms of distance and
column density (namely LSS$\,$1274, HD$\,$191877, and
HD$\,$195965). Their D/O ratios are low, with a weighted average value
of D/O$\;= (1.50\pm0.25) \times 10^{-2}$.  This is 2.5 times lower
(shifted by $9\,\sigma$) than the local value, D/O$\;= (3.84\pm0.16)
\times 10^{-2}$. This result is based on only 3 points, and must be
accepted with caution. However, an examination of
Table~\ref{table_results} shows that most of the targets with
distances approximately $\ge200$~pc have low values of D/O,
Feige$\,$110 being the primary exception. In addition, the three
targets discussed above, all more distant than $\sim500$~pc and with
\ion{D}{1} column densities larger than $7\times10^{15}$~\cmmd, are in
different directions (see Table~\ref{table_results}) and hence, probe
different media. This result suggests the present-epoch D/O ratio is
lower than $3.84\times 10^{-2}$.  The reason for a high local value is
uncertain, but the possibility of unknown deuterium enrichment
process(es) must be considered. A detailed discussion of such
enrichment mechanisms is outside the scope of this report.

It could be argued that the low values of D/O are due to undetected
saturated components. In the case of HD$\,$195965 and HD$\,$191877
Hoopes et al.~(\citealp{hoopes03}) found that high-resolution ground
based \ion{K}{1} data suggested that there were no narrow unresolved
features on either sight line and when they used their STIS data for
HD$\,$195965, they were able to show this more conclusively for that
sight line.  Similar data does not exist for LSS$\,$1274.  Generally
the heavier \ion{O}{1} absorption features should saturate before the
\ion{D}{1} features, 
as at least a part of the line broadening is thermal.
Thus, one would expect that on average, such
effects should tend towards increased values rather than decreased
values as is reported here.
$\gamma\,$Cas, $\epsilon\,$Ori, and $\iota\,$Ori may be exceptions 
to this argument in that $N$(\ion{D}{1}) was measured by 
{\it Copernicus}, but the HST determination of $N$(\ion{O}{1}) 
used the weak intercombination transition at $\lambda\,1356\,$\AA.

\subsection{Local D/H value inferred from D/O}
\label{dsh_value_inferred_from_dso}

The accurate D/O ratio determined for the Local Bubble can be used to
estimate D/H in the Local Bubble, if the value of O/H is known. By
assuming the Meyer et al.~(\citealp{meyer98}) value as updated by
Meyer~(\citealp{meyer01}), O/H$\;=(3.43\pm0.15)\times 10^{-4}$, we
obtain \dshlism$\;=(1.32\pm0.08)\times 10^{-5}$ ($1\,\sigma$). This
value is about 15~\% 
lower than the average values
\dshlism$\;=(1.5\pm0.1)\times10^{-5}$ ($1\,\sigma$) obtained by
Linsky~(\citealp{linsky98}) from the comparison of HST data for 12
nearby sight lines, and
\dshlism$\;=(1.52\pm0.08)\times10^{-5}$ ($1\,\sigma$)
obtained by Moos et al.~(\citealp{moos02}) for the first five lines of
sight observed with \fuse. Although the difference with these direct
measurements is about $2\,\sigma$, which is marginally significant, it
is reassuring that D/H derived from direct and indirect measurements
are so close, lending assurance that the values presently used for the
Local Bubble are at least approximately correct.

What is the source of the disagreement between the two values?  It is
possible that at least one of the three measurements, the D/O (this
study), the D/H (Linsky~\citealp{linsky98}; Moos et
al.~\citealp{moos02}), and the O/H (Meyer et al.~\citealp{meyer98}),
is slightly under- or overestimated.

In a similar manner, using the Meyer et al.~(\citealp{meyer98}) O/H
value, the Linsky~(\citealp{linsky98}) and Moos et
al.~(\citealp{moos02}) D/H values predicts D/O$\;=(4.4\pm0.3)\times
10^{-2}$ ($1\,\sigma$).  The deduced value is high by about
$2\,\sigma$.

This disagreement could be due to the O/H value of Meyer et
al.~(\citealp{meyer98}), which was obtained from GHRS observations of
13 lines of sight ranging from 130 to 1500~pc, most of them being
closer than 500~pc. The Meyer et al. study deals with interstellar
clouds that are more distant than the upper limit we found for the
extent of the region in which D/O was homogeneous:
$\sim150$~pc. However, the O/H value given by Meyer et
al.~(\citealp{meyer98}) agrees with that measured in the more local
ISM by Moos et al.~(\citealp{moos02}) and Oliveira et
al.~(\citealp{oliveira03}) with \fuse. In order for both D/O and D/H
to be respectively $3.8 \times 10^{-2}$ and $1.5 \times 10^{-5}$, the
value of O/H in the Local Bubble would have to be as high as $3.9
\times 10^{-4}$. A local enrichment of oxygen in the interstellar gas 
could exist if the depletion onto dust grains is lower (see
Sect.~\ref{depletion}). However, as discussed above, measurements of
O/H for nearby gas do not show any evidence for such an enrichment.
Note that systematic errors at the 15~\% 
level in the atomic data cannot be ruled out. The Meyer et
al. study used one \ion{O}{1} transition ($\lambda\,1356\,$\AA)
whereas the \fuse\ measurements for the Local Bubble used transitions
in the \fuse\ wavelength range. The two sets of oscillator strengths
have not been compared with sufficient precision to eliminate this
possibility.

Finally, one might ask if the disagreement could be caused by slight
inaccuracies in the local D/H values caused by unknown systematic
errors in the determination of the \ion{H}{1} column densities.
Vidal-Madjar \& Ferlet~(\citealp{avm02}) suggested that direct
measurements of D/H in the very local ISM 
($\log N$(\ion{H}{1})$<10^{19}$~\cmmd) could be 
underestimated by as much as $\sim20$~\%. 
They suggested broad weak components in the interstellar gas that
could lead to {\it overestimates} of $N$(\ion{H}{1}), but would be
undetectable for the much lower \ion{D}{1} column densities.  However,
the Local Bubble D/H ratio deduced here from D/O is lower than the
direct measurements. This suggests that this effect is not prevalent
for most sight lines in the Local Bubble, although the possibility
that it exists for a few sight lines cannot be ruled out.

\subsection{Evidence for a low present-epoch D/H}
\label{present_epoch_dsh}

In this study, a D/H ratio of $(1.32\pm0.08)\times 10^{-5}$ was
estimated for the Local Bubble, and variations start to occur beyond
$\sim150$~pc. At larger distances, there appears to be a downward
trend in the values of D/O.  We reported above that the weighted
average for the three farthest targets of our sample yields a low D/O
ratio, $(1.50\pm0.25) \times 10^{-2}$.  Using the Meyer et
al.~(\citealp{meyer98}) O/H ratio, this D/O translates into
D/H$\;=(5.2\pm0.9)\times 10^{-6}$ ($1\,\sigma$).  
This value is significantly lower than the ones usually assumed, 
by a factor much larger than the 15~\% 
shift found for the Local Bubble in the comparison above.  Although
local material has a higher D/H ratio, raising the sight line
integrated values, the effect will be slight because the local column
densities are so much smaller.
It is possible that the hint of anti-correlation between D/H and O/H,
which was claimed by Steigman~(\citealp{steigman02}) from a study of
the first \fuse\ results presented by Moos et al.~(\citealp{moos02}),
would yield slightly higher O/H ratios on average for these distant
sight lines, and hence, would increase the value of D/H.  However, as
noted above, O/H is generally homogeneous over the distances covered
in this study and if there is an increase, it likely is not strong
enough to increase the value deduced for distant gas to the local D/H
ratio.

While high values of D/H (above $2 \times 10^{-5}$) have been reported
toward two distant targets, namely $\gamma^2\,$Vel (Sonneborn et
al.~\citealp{sonneborn00}) and Feige$\,$110 (Friedman et
al.~\citealp{friedman02}), five low values of D/H (below $1 \times
10^{-5}$) were also previously reported toward distant targets with
{\it Copernicus}, {\it IMAPS}, or \fuse: $\delta\,$Ori$\,$A (Laurent
et al.~\citealp{laurent79}; Jenkins et al.~\citealp{jenkins99}),
$\lambda\,$Sco (York~\citealp{york83}), $\theta\,$Car (Allen et
al.~\citealp{allen92}), and HD$\,$191877 and HD$\,$195965 (Hoopes et
al.~\citealp{hoopes03}).  Values of D/H below $1 \times 10^{-5}$ were
also measured in two Orion regions through ISO observation of HD
(Bertoldi et al.~\citealp{bertoldi99}; Wright et
al.~\citealp{wright99}).  If the present-epoch D/H is actually below
$1 \times 10^{-5}$, as is suggested here, it might require high
deuterium destruction factors for astration processes (see
Sect.~\ref{astration}).

\subsection{D/N ratio: ionization}

Whereas D/O is homogeneous for distances less than 150~pc and
\ion{D}{1} column densities lower than $1 \times 10^{15}$~\cmmd, 
the D/N ratio varies significantly in the near local interstellar
medium. The difference is likely due to the fact that for sufficiently
low \ion{H}{1} column densities, \ion{O}{1} is a better tracer of
\ion{H}{1} than \ion{N}{1} because of ionization effects.  
G191$-$B2B is an example of a local line of sight where nitrogen is
not a good hydrogen tracer (Vidal-Madjar et al.~\citealp{avm98}).
Nitrogen is likely to be more ionized than hydrogen, whereas oxygen
and hydrogen are more strongly coupled to each other by charge
exchange reactions (Sofia \& Jenkins~\citealp{sofia98}; Jenkins et
al.~\citealp{jenkins00}). If ordinary recombination with free
electrons dominates over the charge exchange reaction of
\ion{N}{2} with \ion{H}{1}, we expect to have 
$n_{\rm N\,II}/n_{\rm N\,I} > 
n_{\rm H\,II}/n_{\rm H\,I}$, 
since the \ion{N}{1} photo-ionization cross section is larger than
that of \ion{H}{1} (H\'ebrard et al.~\citealp{hebrard02}).  This
inequality with the hydrogen ionization fraction does not arise for
oxygen because the charge exchange cross section of \ion{O}{2} with
\ion{H}{1} is considerably larger than for \ion{N}{2} with
\ion{H}{1}. Thus, as a substantial fraction of the LISM is partially 
ionized in the Local Bubble, $N$(\ion{D}{1})/$N$(\ion{N}{1}) is not a
good approximation for D/N. 
Indeed, \ion{N}{2} has abundances similar to those of \ion{N}{1} 
in the Local Bubble (see, \ie, Wood et al.~\citealp{wood03}) and there 
are even at least three targets for which $N$(\ion{N}{2})$\;>N$(\ion{N}{1}): 
HZ$\,$43$\,$A (Kruk et al.~\citealp{kruk02}), WD$\,$1634$-$573 
(Wood et al.~\citealp{wood02}), and WD$\,$2211$-$495 
(H\'ebrard et al.~\citealp{hebrard02}).

In addition, if
$N$(\ion{N}{2})/$N$(\ion{N}{1}) varies because of an inhomogeneous
flux of EUV photons, the $N$(\ion{D}{1})/$N$(\ion{N}{1}) ratio may
also vary, being higher toward the most ionized regions. Tests on
species highly dependent on ionization, like \ion{Ar}{1}, or on ratios
such as $N$(\ion{Si}{2})/$N$(\ion{Si}{1}), may help to determine if
there is a correlation between ionization and the D/N variations which
we detected. This is beyond the scope of the present work on D/O. For
a study of ionization conditions in the Local Bubble and the LISM
slightly beyond, see Slavin \& Frisch~(\citealp{slavin02}) 
or Lehner et al.~(\citealp{lehner03}). 

The problem of nitrogen ionization is less critical outside the 
Local Bubble, where the N/H ratio has been shown to be homogeneous 
for \ion{H}{1} column densities $\ge1\times 10^{20}$~\cmmd\ 
(Ferlet~\citealp{ferlet81}; Meyer et al.~\citealp{meyer97}). 

\begin{figure}
\psfig{file=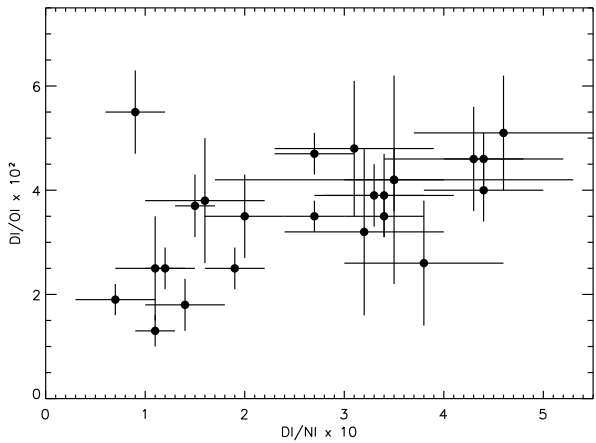,height=6.5cm}
\caption{D/O measurements in the interstellar medium 
as a function of D/N.  
The plotted error bars are~$1\,\sigma$.
\label{fig_dsn_dso}}
\end{figure}

For high column densities, the D/N ratio may be a valuable D/H proxy,
although it will not be as secure as D/O until more extensive studies
of the chemical abundances of \ion{N}{1} in the ISM are
performed. Figure~\ref{fig_dsn_dso} shows the D/O as a function of the
D/N ratio for our sample. A correlation seems to appear, but with
large scatter; the lowest D/N ratios are generally measured toward
targets with low D/O ratios. These are also the more distant targets.
The one exception is Feige$\,$110 with a high D/O and a highly
uncertain distance between 112 and 444~pc.  This correlation can be
understood in part by noting that at low column densities, the D/O
ratio is higher than average and the D/N ratio is also higher due to
ionization effects, whereas at high column densities, we have found
D/O to be lower and D/N is also expected to decrease as the effects of
EUV ionization falls off.

The weighted average of D/N for the three farthest targets
(LSS$\,$1274, HD$\,$191877, and HD$\,$195965) is D/N$\;= (1.15\pm0.16)
\times 10^{-1}$, without significant variations. Assuming the N/H
ratio measured by Meyer et al.~(\citealp{meyer97}), N/H$\;=(7.5
\pm0.4) \times 10^{-5}$, our D/N ratio for distant gas yields
D/H$\;=(8.6\pm1.3)\times 10^{-6}$ ($1\,\sigma$). This is not in
perfect agreement with the D/H ratio obtained through D/O; the error
bars overlap at the $2\,\sigma$ level. This could be caused by
ionization and/or variations in the chemical abundance. In any case,
this value is significantly lower than the D/H value usually assumed
for the present epoch.

\subsection{Oxygen depletion}
\label{depletion}

Whereas the \ion{D}{1} column density represents the total amount of
neutral deuterium in the interstellar medium observed here, this
is not the case for \ion{O}{1}. A significant fraction of the
interstellar oxygen is depleted onto dust grains, and this amount does
not contribute to the measured absorption.  Cardelli et
al.~(\citealp{cardelli96}) estimated the relative amount of oxygen in
the solid phase to be O/H$\,\simeq1.8\times 10^{-4}$. 
Thus, in addition to the gaseous oxygen measured by Meyer et
al.~(\citealp{meyer98}), O/H$\,\simeq3.4\times 10^{-4}$, the total
interstellar oxygen should be O/H$\,\simeq5.2\times 10^{-4}$.
This value agrees with the Solar value obtained by Allende Prieto et
al.~(\citealp{allende01}): O/H$\,=(4.9\pm0.6)\times 10^{-4}$ (a lower
value than the ones previously reported). Thus, about one third of the
total oxygen is depleted onto dust grains.

When using our D/O measurement, one considers only the gas phase
abundance of the neutral oxygen. However, the same holds when using
the O/H interstellar measurement, so no depletion corrections need to
be applied in order to translate a D/O result into a D/H
measurement. More importantly, studies by Meyer et
al.~(\citealp{meyer98}), Cartledge et al.~(\citealp{cartledge01}), and
Andr\'e et al.~(\citealp{andre03}) have shown that the O/H ratio does
not vary and hence, oxygen depletion is constant out to distances of
more than 1~kpc, over a wide range of fractional H$_2$ abundances
and \ion{H}{1} column densities.  Cartledge et
al.~(\citealp{cartledge01}) found lower oxygen abundances in a few
translucent clouds on the line of sight to distant stars. This first
hint of oxygen depletion enhancement should imply an increased value
of the D/O for sight lines of this type. However, those lines of sight
were much denser than the ones reported here, in which molecular
hydrogen is at most a few percent, and generally less than $10^{-4}$
of the \ion{H}{1} column.

\subsection{Deuterium evolution}
\label{astration}

There are primarily two ways to examine the evolution of the abundance
of deuterium throughout the history of the Universe: first, from
Galactic abundance gradients, and second, from abundances at different
stages of evolution (\ie\ studies of the interstellar {\it vs.}
intergalactic media).

The Galactic abundance gradients are obtained in models of 
chemical evolution, assuming a radially dependent star formation rate
and a timescale for formation of the disk by infall that 
depends on the galactocentric distance (see, \eg, Hou et 
al.~\citealp{hou00}).
Inside-out formation of the Galaxy is also suggested by more
sophisticated dynamical models (see, \eg, Samland \& 
Gerhard~\citealp{samland03}).
As astration destroys deuterium and produces oxygen, the deuterium 
abundance should be lower in the inner part of the Galaxy compared 
to the outer part, and oxygen more abundant in the inner part than 
in the outer. Thus, the D/O ratio should have an even steeper 
gradient, from low values in the center to high values in the outer 
part of the Galaxy. 

From their ``two-infall model'', Chiappini et 
al.~(\citealp{chiappini02}) predict the galactocentric gradient 
$d\log ({\rm D/O})/dR\simeq0.13$~dex/kpc. Assuming that the local 
value, D/O$\;= (3.84\pm0.16) \times 10^{-2}$, reported here is typical 
of this region of the Galaxy implies D/O ratios 
$\sim2.8\times 10^{-2}$ and 
$\sim5.2\times 10^{-2}$ should be observed in 1~kpc distant clouds 
toward respectively the center and the anticenter of the Galaxy. 
Given our typical error bars, such gradients are 
detectable with \fuse. They should appear as a correlation of the D/O 
ratio with the Galactic longitude, especially for the distant targets. 
Up to now, we have not detected such a signature, but our sample of 
distant targets is still poor: 
only five targets could be more distant than 500~pc. 

We note that Chiappini et al.~(\citealp{chiappini02}) predict a D/O
value between $\sim2.5\times 10^{-2}$ and $\sim4.0\times 10^{-2}$ at a
galactocentric distance of 8~kpc, \ie\ the distance of the LISM.
Because the value predicted by Chiappini et
al.~(\citealp{chiappini02}) refers to all the oxygen,
\ie\ even that which is depleted on the dust grains, it is
necessary to decrease the value for the Local Bubble,
D/O$\;=(3.84\pm0.16) \times 10^{-2}$, by about 30~\%.
The total local D/O is then around $2.7 \times 10^{-2}$, 
in agreement with
the value predicted by Chiappini et al.~(\citealp{chiappini02}).  
Note that the value of D/H in the Local Bubble may be high 
compared to other regions, as discussed above. 

Abundance measurements at different stages of evolution are the second
way to constrain the evolution of deuterium.  Whereas the abundances
in the interstellar medium are representative of the present epoch,
primordial abundances likely exist in the intergalactic
low-metallicity clouds observed towards quasars at high
redshifts. These values for the primordial abundance
may also be compared with those deduced
independently from the cosmic microwave background (CMB) anisotropy
studies (see, \eg, Cyburt et al.~\citealp{cyburt03}; Spergel et
al.~\citealp{spergel03}). 
Primordial D/H values between
$2.5$ and $3.0\times10^{-5}$ are the preferred ones, as they are
observed toward several quasars (O'Meara et al.~\citealp{omeara01};
Kirkman et al.~\citealp{kirkman03}) and they are in quite good
agreement with CMB measurements performed with different missions such
as BOOMERANG (de~Bernardis et al.~\citealp{bernardis02}), DASI (Pryke
et al.~\citealp{pryke02}), ARCHEOPS (Beno\^{\i}t et
al.~\citealp{benoit03}), or now WMAP (Bennett et
al.~\citealp{bennett03}). These measurements imply a baryon-to-photon
ratio around $\eta\sim6\times10^{-10}$.  Note however, the different
results from MAXIMA (Stompor et al.~\citealp{stompor01}), the high D/H
measurement reported by Webb et al.~(\citealp{webb97}), and the fact
that the baryonic density inferred from the low D/H values through BBN
is not in good agreement with those based on helium and lithium
primordial abundances.

Standard chemical evolution models predict a D/H decrease by a factor
2 to 3 in 15~Gyrs (\eg,~Galli et al.~\citealp{galli95};
Prantzos~\citealp{prantzos96}; Tosi et al.~\citealp{tosi98}).  More
recent models predict smaller decreases, by factors $\sim1.5$
(Chiappini et al.~\citealp{chiappini02}). With such low factors, it
becomes more difficult to link the primordial D/H measurements, found to 
be between $2.5$ and $3.0\times10^{-5}$, with the one we deduced here for
the local interstellar medium, around $1.3\times10^{-5}$. The problem
becomes more severe if the present-epoch D/H is below
$1\times10^{-5}$, as suggested above.  Note however that there are
some nonstandard models which propose higher astration ratios
(Vangioni-Flam \& Cass\'e~\citealp{flam95}; Scully et
al.~\citealp{scully97}).

In addition, the estimates of the proto-solar abundances of deuterium
favor values of D/H around $2-3\times10^{-5}$ (Gautier \& 
Morel~\citealp{gautier97}; Mahaffy et al.~\citealp{mahaffy98};
Lellouch et al.~\citealp{lellouch01}, Hersant 
et al.~\citealp{hersant01}), \ie\ near the primordial ones.
This result allows only low astration factors during the first ten
Gyears, most of the astration occurs in the recent Universe. Such
assumptions jeopardize standard models.

Up to now, only a few measurements of deuterium in primordial media 
have been made. In QSO$\,$1937$-$1009 and QSO$\,$1009+2956, Burles \& 
Tytler~(\citealp{bt1}; \citealp{bt2}) reported \ion{D}{1} detections
which allow measurements of the primordial D/H, but they obtained only
upper limits on the \ion{O}{1} column densities.  O'Meara et
al.~(\citealp{omeara01}) presented the first simultaneous measurements
of \ion{D}{1} and \ion{O}{1} column densities toward the $z\simeq2.5$
Lyman limit system on the line of sight to QSO$\,$0105+1619 and
obtained D/O$\;=(280\pm30)\times10^{-2}$.  Levshakov et
al.~(\citealp{levshakov02}) reported also an
\ion{O}{1} detection in the $z\simeq3.0$ damped \lya\ system toward
QSO$\,$0347$-$3819, in which D'Odorico et al.~(\citealp{dodorico01})
previously reported a \ion{D}{1} measurement. That new study yields 
the following value: D/O$\;=(37\pm3)\times 10^{-2}$. Note
that when using the initial D'Odorico et al.~(\citealp{dodorico01})
$N$(\ion{D}{1}) measurement, that ratio is reduced to
D/O$\;=(21\pm4)\times 10^{-2}$.
Finally, Kirkman et al.~(\citealp{kirkman03}) recently detected both 
\ion{D}{1} and \ion{O}{1} in a third intergalactic cloud, 
at a redshift $z\simeq2.5$ toward the quasar 
QSO$\,$1243+3047. They found a high 
ratio: D/O$\;=(3000\pm300)\times 10^{-2}$ 
(all the error bars given here are within a \1s\ confidence interval).
Thus, the few primordial D/O measurements available are significantly
higher than our local interstellar D/O value. The local D/O is 
decreased by factors between $\sim5$ and $\sim70$ with respect to the
primordial ones, and even by a factor as high as $\sim800$ when
compared to the high D/O value reported by Kirkman et
al.~(\citealp{kirkman03}).  This result confirms that D/O is a very
sensitive probe of astration. According to Chiappini et
al.~(\citealp{chiappini02}), the D/O values predicted in the outer
parts of the Galaxy should be close to the primordial value. From
their model, they computed D/O values between $\sim25\times 10^{-2}$
and $\sim100\times 10^{-2}$ for galactocentric distances between 15
and 18~kpc, respectively.

\section{Conclusion}

We have presented the first results of a \fuse\ survey of D/O in the
interstellar medium. D/O is a good D/H proxy because it does not
require ionization or significant depletion corrections.  Using oxygen
instead of hydrogen also reduces the systematic errors for the
measured column densities.  Only six D/O measurements were available
prior to the \fuse\ mission; the present sample totals 24 lines of
sight.

We find that the D/O ratio is homogeneous in the near local 
interstellar medium, with a weighted mean  
D/O$\;= (3.84\pm0.16) \times 10^{-2}$ ($\sim4$~\% 
uncertainty in the mean at \1s). This value is representative 
of the interstellar medium in the Local Bubble, for distances 
$<150$~pc and $\log N($\ion{D}{1}$) \leq 14.5$. 
The remarkable homogeneity of D/O  is a strong argument for 
both D/H and O/H stability at the scale of the Local Bubble. 

The local D/O value and the average value of O/H in the local
interstellar medium obtained by Meyer et al.~(\citealp{meyer98}) imply
that \dshlism$\;=(1.32\pm0.08)\times 10^{-5}$. This value is slightly
lower than the values obtained from direct D/H measurements in the
Local Bubble ($(1.52\pm0.08)\times10^{-5}$; Moos et
al.~\citealp{moos02}) or the Local Interstellar Cloud
($(1.5\pm0.1)\times10^{-5}$; Linsky~\citealp{linsky98}).  Linking
these local values with the primordial ones is marginal if astration
factors around 1.5 (Chiappini et al.~\citealp{chiappini02}) are
used. 
Using a low deuterium astration factor of 1.5, our
local D/H value translates into a primordial value around
$2\times10^{-5}$.  According to the standard Big Bang Nucleosynthesis
predictions, this corresponds to a baryon-to-photon ratio around
$\eta=7.2\times10^{-10}$, slightly higher than most of the
determinations made from the anisotropy of the cosmic microwave
background. 

Beyond the Local Bubble, we found significant variations in the value
of D/O, implying D/H variations.  In addition, the D/O values are
typically lower.  Indeed, the more distant targets suggest a 
present-epoch deuterium abundance below
$1\times 10^{-5}$, \ie\ lower than the
value usually assumed.  As these sight lines are in different
directions on the sky, it follows that the local value of D/H may be
high compared to other nearby regions in the Galaxy. 

We did not detect any correlation between the target-to-target
variations and the changes with galactic longitude expected to result
from predicted galactocentric abundance gradients. However, the number
of distant targets is quite small and additional D/O studies of
distant lines of sight will be needed to detect such gradients.
Together with $N($\ion{H}{1}$)$ estimations, additional measurements
would also shed light on the hint of anti-correlation between D/H and
O/H reported by Steigman~(\citealp{steigman02}).  Finally, additional
studies of distant sight lines will confirm or disprove the evidence
for low deuterium abundance in the present epoch implied by the date
reported here.  If the low D/O values measured at large distances are
representative of the present-epoch value, rather than the local one,
this would further increase the difficulty in linking the present and
primordial deuterium abundance values through astration correction.

\acknowledgments
This work is based on data obtained by the NASA-CNES-CSA \fuse\ 
mission operated by the Johns Hopkins University. Financial support to
H.W.M. was provided by NASA contract NAS5-32985. G.H. was supported by
CNES.  This work used the profile fitting procedure Owens.f developed
by M. Lemoine and the French \fuse\ Team.  We would like to thank
D. C. Morton for useful comments about atomic data, J.-M. D\'esert,
S. Lacour, and A. Moullet for their help in data processing, and
E. Vangioni-Flam and S. Boissier for useful discussions.

This study is part of a extensive program in which all of the D/H
Working Group of the FUSE Principal Investigator Team participated.
We especially acknowledge M. K. Andr\'e, W. P. Blair, P. Chayer,
J. Dupuis, R. Ferlet, S. D. Friedman, C. Gry, C. G. Hoopes, J. C. Howk,
E. B. Jenkins, D. C. Knauth, J. W. Kruk, A. Lecavelier des \'Etangs,
N. Lehner, M. Lemoine, J. L. Linsky, C. M. Oliveira, K. R. Sembach, 
J. M. Shull, G. Sonneborn, P. Sonnentrucker, T. M. Tripp, 
A. Vidal-Madjar, G. M. Williger, B. E. Wood, and D. G. York. 
We thank them for many discussions and comments with respect 
to the deuterium studies.

%%%%%%%%%%%%%%%%%%%%%%%%%%%%%%%%%%%%%%%%%%%%%%%%%%%%%%%%%%%%%%%%%%%

%%%%%%%%%%%%%%%%%%%%%%%%%%%%%%%%%%%%%%%%%%%%%%%%%%%%%%%%%%%%%%%%%%%

%%%%%%%%%%%%%%%%%%%%%%%%%%%%%%%%%%%%%%%%%%%%%%%%%%%%%%%%%%%%%%%%%%%

\newpage

\begin{table*}
\caption{Log of the observations.} 
\begin{tabular}{lccrrccc}
\hline
\hline
Target & Observation date & Observation reference & $T_{\rm 
obs}\,({\rm ksec})^{\rm a}$ 
\hspace{-0.8cm} & \ \ \ 
$N_{\rm exp}^{\rm b}$ & Aperture$^{\rm c}$ & Mode$^{\rm d}$ & 
CalFUSE$^{\rm e}$ \\
\hline
Sirius~B & 2001 Nov 25 & C1160102 & 1.6 & 4 & MDRS & HIST & 1.8.7 \\
         & 2001 Nov 17 & C1160103 & 1.3 & 1 & MDRS & TTAG & 1.8.7 \\
         & 2001 Nov 25 & C1160104 & 1.8 & 4 & MDRS & HIST & 1.8.7 \\
       \multicolumn{3}{r}{Total:} & 4.7 & 8 \\
\hline
WD$\,$2004$-$605 & 2001 May 21 & P2042201 & 10.2 & 12 & MDRS & TTAG & 
2.1.6 \\
                 & 2001 Aug 31 & P2042202 & 26.0 & 28 & MDRS & TTAG & 
2.1.6 \\
                 & 2002 Apr 10 & P2042203 & 37.6 & 37 & MDRS & TTAG & 
2.1.6 \\
               \multicolumn{3}{r}{Total:} & 73.8 & 77 \\
\hline
WD$\,$1631$-$781 & 2000 Jan 18 & P1042901 & 22.1 &  9 & MDRS & TTAG & 
2.0.4 \\
                 & 2001 Jan 31 & P1042902 & 30.2 & 33 & MDRS & TTAG & 
2.0.4 \\
               \multicolumn{3}{r}{Total:} & 52.3 & 42 \\
\hline
CPD$-$31$\,$1701 & 2001 Apr 06 & P2050601 &  7.2 & 12 & LWRS & HIST & 
2.0.5 \\
                 & 2001 Nov 01 & P2050602 &  9.3 & 19 & MDRS & HIST & 
2.0.5 \\
               \multicolumn{3}{r}{Total:} & 16.5 & 31 \\
\hline
BD$\,+28^{\circ}\,4211$
          & 2000 Jun 13 & M1080901 &  2.2 &  4 & LWRS & HIST & 2.1.1 \\
          & 2000 Jul 16 & M1040101 & 16.7 & 36 & LWRS & HIST & 2.1.0 \\
          & 2000 Sep 19 & M1040105 &  7.9 & 17 & LWRS & HIST & 2.1.1 \\
          & 2001 Jul 29 & M1031201 &  3.0 &  5 & LWRS & HIST & 2.1.0 \\
          & 2001 Jul 31 & M1031204 &  2.9 &  5 & LWRS & HIST & 2.1.0 \\
          & 2000 Jul 17 & M1040102 & 24.8 & 52 & MDRS & HIST & 2.1.0 \\
          & 2001 Jul 29 & M1031202 &  2.9 &  6 & MDRS & HIST & 2.1.0 \\
          & 2001 Jul 31 & M1031205 &  2.9 &  6 & MDRS & HIST & 2.1.0 \\
          & 2001 Jul 29 & M1031203 &  2.4 &  5 & HIRS & HIST & 2.1.0 \\
          & 2001 Jul 31 & M1031206 &  2.4 &  5 & HIRS & HIST & 2.1.0 \\
        \multicolumn{3}{r}{Total:} & 68.1 & 141 \\
\hline
LSS$\,$1274
          & 2002 Mar 08 & P2051702 &  8.0 & 11 & MDRS & TTAG & 2.1.6 \\
          & 2002 Mar 11 & P2051701 & 14.0 & 12 & MDRS & TTAG & 2.1.6 \\
          & 2002 May 02 & P2051703 & 63.0 & 54 & MDRS & TTAG & 2.1.6 \\
        \multicolumn{3}{r}{Total:} & 85.0 & 77 \\
\hline
Feige$\,$110
          & 2000 Jun 22 & M1080801 &  6.2 &  8 & LWRS & HIST & 1.8.7 \\
          & 2000 Jun 30 & P1044301 & 21.8 & 48 & LWRS & HIST & 1.8.7 \\
        \multicolumn{3}{r}{Total:} & 28.0 & 56 \\
\hline
\end{tabular}
\label{table_obslog}
\\
$^{\rm a}$ Total exposure time of the observation (in $10^3$~s). \\
$^{\rm b}$ Number of individual exposures during the observation. \\
$^{\rm c}$ LWRS, MDRS, and HIRS are, respectively, large, medium, 
and narrow \fuse\ slits. \\
$^{\rm d}$ HIST and TTAG are respectively histogram and time-tagged 
photon 
address \fuse\ modes. \\
$^{\rm e}$ Version of the pipeline used for spectral extraction. \\
\end{table*}

\begin{table}
\caption{\ion{D}{1}, \ion{O}{1}, and \ion{N}{1} lines included 
in the final fits.} 
\begin{tabular}{ccccccccc}
\hline
\hline
 & & 
\begin{rotate}{Sirius~B}\end{rotate} & 
  \begin{rotate}{WD$\,$2004$-$605}\end{rotate} &
    \begin{rotate}{WD$\,$1631$+$781}\end{rotate} & 
      \begin{rotate}{CPD$-$31$\,$1701}\end{rotate} & 
        \begin{rotate}{BD$\,+28^{\circ}\,4211$}\end{rotate} &
          \begin{rotate}{LSS$\,$1274}\end{rotate} &
            \begin{rotate}{Feige$\,$110}\end{rotate} \\
\hline
$\lambda^{\rm a}$ & $f^{\rm b}$ & \#$^{\rm c}$ & \#$^{\rm c}$ 
& \#$^{\rm c}$ & \#$^{\rm c}$ & \#$^{\rm c}$ & \#$^{\rm c}$ & 
\#$^{\rm c}$ \\
\hline
\multicolumn{2}{c}{\ion{D}{1}} \\
972.2723     & $2.90\times10^{-2}$ & 
  &
  2 & 
    2 & 
        & 
          &
            \\
949.4848     & $1.39\times10^{-2}$ & 
  &
  2 & 
    1 & 
      4 & 
          &
            \\
937.5484     & $7.80\times10^{-3}$ & 
  &
  2 &
    1 & 
      4 & 
        3 & 
            \\
930.4951     & $4.82\times10^{-3}$ & 
  &
  2 & 
    2 & 
        &
        3 &
           \\
925.9738     & $3.18\times10^{-3}$ & 
  &
  2 & 
    2 & 
      4 & 
        4 & 
           \\
922.8993     & $2.22\times10^{-3}$ & 
  &
  2 & 
    2 & 
        &
        3 &
            \\
920.7126     & $1.61\times10^{-3}$ & 
  &
  2 & 
    1 & 
      3 & 
        6 &
          2  \\
919.1013     & $1.20\times10^{-3}$ & 
  &
    & 
      & 
        & 
        1 &
          2 \\
917.8797     & $9.21\times10^{-4}$ & 
  &
    & 
      & 
        & 
        2 & 
          2 \\
916.9311     & $7.23\times10^{-4}$ & 
  &
  1 & 
    1 & 
        & 
          & 
            \\
\multicolumn{2}{r}{Total:} & 
   & 
  15 & 
    12 & 
      15 & 
        22 &
           6 &
              -- \\
\hline
\multicolumn{2}{c}{\ion{O}{1}} \\
988.7734     & $4.65\times10^{-2}$ & 
2 \\
971.738$^d$  & $1.38\times10^{-2}$ & 
1 \\
1039.2304    & $9.20\times10^{-3}$ & 
3 \\
988.6549     & $8.30\times10^{-3}$ & 
2 \\
950.8846     & $1.58\times10^{-3}$ & 
  & 
  2 & 
      &
        &
          &
            \\
921.8570$^{\rm d}$ & $1.19\times10^{-3}$ & 
  &
  2 & 
      & 
      4 & 
          &
            \\
919.6580$^{\rm d}$ & $9.47\times10^{-4}$ & 
  &
  2 & 
      & 
      4 & 
          &
           \\
988.5778     & $5.53\times10^{-4}$ & 
2 \\
930.2566     & $5.37\times10^{-4}$ & 
  &
  2 & 
    2 & 
      4 & 
          &
            \\
916.8150$^{\rm d}$ & $4.74\times10^{-4}$ & 
  &
  1 & 
    1 &
        & 
          &
            \\
925.4460 & $3.50\times10^{-4}$ & 
  &
  2 & 
    2 & 
      1 &
        3 &
           \\
922.2000     & $2.45\times10^{-4}$ & 
  &
  2 & 
    2 & 
      2 & 
        1 & 
           \\
919.9170     & $1.78\times10^{-4}$ & 
  &
  2 & 
    1 & 
      4 & 
        6 &
           \\
974.0700     & $1.56\times10^{-5}$ & 
  &
    & 
      & 
        &
        4 &
           2 \\
\multicolumn{2}{r}{Total:} & 
10 &
  15 &
     8 & 
      19 & 
        14 &
           2 &
             -- \\
\hline
\multicolumn{2}{c}{\ion{N}{1}} \\
964.6256     & $9.43\times10^{-3}$ & 
  &
  1 & 
      &
        &
          &
            \\
954.1042     & $6.76\times10^{-3}$ & 
  &
  2 & 
      &
        &
          & 
           \\
965.0413     & $4.02\times10^{-3}$ & 
  &
  1 &
      & 
        & 
          & 
           \\
952.3034     & $1.87\times10^{-3}$ & 
  &
  2 & 
    2 & 
      4 &
        3 & 
           \\
952.4148     & $1.70\times10^{-3}$ & 
  &
  2 & 
    2 & 
      4 &
        3 & 
          \\
952.5227     & $6.00\times10^{-4}$ & 
  &
  2 & 
    2 & 
      4 &
        3 & 
           \\
951.0791     & $1.29\times10^{-4}$ & 
  &
    & 
      & 
        &
        4 & 
          2 &
            2 \\
955.8816     & $5.88\times10^{-5}$ & 
  &
    & 
      & 
        &
        2 & 
           1 \\
955.2644     & $5.03\times10^{-5}$ & 
  &
    & 
      & 
        &
        3 & 
           \\
959.4937     & $4.98\times10^{-5}$ & 
  &
    & 
      & 
        &
          &
           2 &
               \\
951.2947     & $1.81\times10^{-5}$ & 
  &
    & 
      & 
        &
        4 &
          2  &
             2 \\
955.5294     & $1.72\times10^{-5}$ & 
  &
    & 
      & 
        &
          &
           1 &
               \\
\multicolumn{2}{r}{Total:} & 
    &
   10 & 
      6 & 
       12 &
         22 &
            8 &
              4 \\
\hline
\end{tabular}
\label{table_lines}
\\
$^{\rm a}$ Wavelength in vacuum at rest (in \AA). \\
$^{\rm b}$ Oscillator strength. \\
$^{\rm c}$ Number of independent lines included in the fit (observed 
on different channels and through different slits, 
using all the observations presented in Table~\ref{table_obslog}). \\
$^{\rm d}$ Triplet structure used 
(see Morton~\citealp{morton91,morton02}) \\
\end{table}

\begin{table*}
\begin{small}
\caption{\ion{D}{1}, \ion{O}{1}, and \ion{N}{1} interstellar column 
densities
measured toward 24 lines of sight.} 
\hspace{-1cm}
\begin{tabular}{lcccccccccl}
\hline
\hline
Target & Sp. & $l\;(^{\circ})$ & 
$b\;(^{\circ})$ & $d$ (pc)$\,^{\rm a}$ 
& $\log N$(\ion{D}{1})$\,^{\rm b}$ 
& $\log N$(\ion{N}{1})$\,^{\rm b}$ 
& $\log N$(\ion{O}{1})$\,^{\rm b}$ 
& D/N $\times10$ & D/O $\times10^2$ & Ref.$\,^{\rm c}$ \\
\hline
Sirius~B    & DA2 &  227.2 & $-8.9$ & $2.64\pm0.01\,^{\rm h}$ 
& $12.88\pm0.08$ 
& $13.35\pm0.03$ 
& $14.29\pm0.05$ & $3.4\pm0.7$ & $3.9\pm0.8$ & 1,2 \\
HZ$\,$43$\,$A    & DA1 &  54.1 & $+84.2$ & $68\pm13\,^{\rm p}$ 
& $13.15\pm0.02$ 
& $13.51\pm0.03$ 
& $14.49\pm0.04$ & $4.4\pm0.4$ & $4.6\pm0.5$ & 3 \\
G191$-$B2B       & DA1 & 155.9 &  $+7.1$ & $69^{+19}_{-12}\,^{\rm h}$ 
& $13.40\pm0.04$ 
& $13.87\pm0.04$ 
& $14.86\pm0.04$ & $3.4\pm0.4$ & $3.5\pm0.4$ & 4 \\ 
Capella     & G8III+ &  162.6 & $+4.6$ & $12.9\pm0.2\,^{\rm h}$ 
& $13.44\pm0.01$ 
& $13.86\pm0.09$ 
& $15.02\pm0.16$ & $3.8\pm0.8$ & $2.6\pm1.2$ & 5 \\
WD$\,$0621$-$376 & DA1 & 245.4 & $-21.4$ & $78\,^{\rm p}$ 
& $13.85\pm0.05$ 
& $14.34\pm0.05$ 
& $15.26\pm0.04$ & $3.3\pm0.5$ & $3.9\pm0.6$ & 6 \\
WD$\,$2211$-$495 & DA1 & 345.8 & $-52.6$ & $53\,^{\rm p}$ 
& $13.94\pm0.05$ 
& $14.30\pm0.03$ 
& $15.34\pm0.04$ & $4.4\pm0.6$ & $4.0\pm0.6$ & 7 \\
WD$\,$1634$-$573 & DO  & 329.9 &  $-7.0$ & $37\pm3\,^{\rm h}$ 
& $14.05\pm0.03$ 
& $14.62\pm0.04$ 
& $15.51\pm0.03$ & $2.7\pm0.3$ & $3.5\pm0.3$ & 8 \\
WD$\,$2331$-$475 & DA1 & 334.8 & $-64.8$ & $82\,^{\rm p}$
& $14.19\pm0.06$ 
& $14.53\pm0.05$ 
& $15.48\pm0.06$ & $4.6\pm0.9$ & $5.1\pm1.1$ & 9 \\
$\alpha\,$Vir & B1III-IV & 316.1 & $+50.8$ & $80\pm6\,^{\rm h}$
& $14.20\pm0.20$ 
& $14.66\pm0.05$ 
& $15.58\pm0.10$ & $3.5\pm1.8$ & $4.2\pm2.0$ & 10 \\
GD$\,$246        & DA1 &  87.2 & $-45.1$ & $79\,^{\rm p}$ 
& $14.29\pm0.05$ 
& $14.75\pm0.03$ 
& $15.67\pm0.04$ & $3.5\pm0.5$ & $4.2\pm0.6$ & 9 \\
WD$\,$2004$-$605 & DA1 & 336.6 & $-32.9$ & $52\,^{\rm p}$ 
& $14.33\pm0.08$ 
& $14.84\pm0.08$ 
& $15.65\pm0.08$ & $3.1\pm0.8$ & $4.8\pm1.3$ & 1 \\
HZ$\,$21        & DO2 &  175.0 & $+80.0$ &  $115^{\rm p}$ 
& $14.40\pm0.08$ 
& $14.77\pm0.04$ 
& $15.74\pm0.05$ & $4.3\pm0.9$ & $4.6\pm1.0$ & 9 \\
WD$\,$1631$+$781 & DA1 & 111.3 & $+33.6$ & $67\,^{\rm p}$ 
& $14.48\pm0.10$ 
& $15.28\pm0.10$ 
& $15.90\pm0.09$ & $1.6\pm0.6$ & $3.8\pm1.2$ & 1 \\
CPD$-$31$\,$1701 & sdO & 246.5 &  $-5.5$ & $131\pm28\,^{\rm h}$ 
& $14.50\pm0.06$ 
& $15.33\pm0.04$ 
& $15.93\pm0.04$ & $1.5\pm0.2$ & $3.7\pm0.6$ & 1 \\
BD$\,+28^{\circ}\,4211$ & sdO & 81.9 & $-19.3$ & $104\pm18\,^{\rm h}$ 
& $14.95\pm0.02$ 
& $15.52\pm0.05$ 
& $16.28\pm0.03$ & $2.7\pm0.4$ & $4.7\pm0.4$ & 1 \\
$\delta\,$Ori$\,$A & O9.5II & 203.9 & $-17.7$ & $280\pm80\,^{\rm h}$
& $15.06\pm0.05$ 
& $15.79\pm0.04$ 
& $16.67\pm0.05$ & $1.9\pm0.3$ & $2.5\pm0.4$ & 11,12 \\
$\gamma\,$Cas & B0IVe & 123.6 & $-2.1$ & $190\pm20\,^{\rm h}$
& $15.15\pm0.05$ 
& $16.06\pm0.04$ 
& $16.76\pm0.04$ & $1.2\pm0.2$ & $2.5\pm0.4$ & 12,13,14 \\
Lan$\,$23 & DA & 107.6 & $-0.6$ & $122\,^{\rm p}$ 
& $15.23\pm0.07$ 
& $15.73\pm0.07$
& $16.72\pm0.17$ & $3.2\pm0.8$ & $3.2\pm1.6$ & 9 \\
$\epsilon\,$Ori & B0Ia & 205.2 & $-17.2$ & $450\pm200\,^{\rm h}$
& $15.26\pm0.05$ 
& $16.45\pm0.20$ 
& $16.98\pm0.05$ & $0.7\pm0.4$ & $1.9\pm0.3$ & 12,15,16 \\
$\iota\,$Ori & O9III & 209.5 & $-19.6$ & $410\pm150\,^{\rm h}$
& $15.30\pm0.05$ 
& $15.99\pm0.07$ 
& $16.76\pm0.09$ & $2.0\pm0.4$ & $3.5\pm0.8$ & 12,14,15 \\
Feige$\,$110 & sdOB & 74.1 & $-59.1$ & $179^{+265}_{-67}\,^{\rm h}$
& $15.47\pm0.03$ 
& $16.51\pm0.10$ 
& $16.73\pm0.05$ & $0.9\pm0.3$ & $5.5\pm0.8$ & 1,17 \\
LSS$\,$1274 & sdO & 277.0 & $-5.3$ & $580\pm100\,^{\rm p}$
& $15.87\pm0.10$ 
& $16.73\pm0.05$ 
& $17.62\pm0.08$ & $1.4\pm0.4$ & $1.8\pm0.5$ & 1 \\
HD$\,$195965 & B0V & 85.7 & $+5.0$ & $800\,^{\rm p}$ 
& $15.88\pm0.07$
& $16.85\pm0.03$
& $17.77\pm0.03$   & $1.1\pm0.2$ & $1.3\pm0.3$ & 18 \\
HD$\,$191877 & B1Ib & 61.6 & $-6.5$ & $2200^{\rm p}$ 
& $15.94\pm0.10$
& $16.88\pm0.05$
& $17.54\pm0.10$ & $1.1\pm0.4$ & $2.5\pm1.0$ & 18 \\
\hline
\end{tabular}
\label{table_results}
\\
$^{\rm a}$ Distances are photometric ($D\,^{\rm p}$; see references) or 
from 
Hipparcos parallax ($D\,^{\rm h}$; Perryman et 
al.~\citealp{perryman97}). \\
$^{\rm b}$ Column densities $N$ (in \cmmd); $1\,\sigma$ error bars. \\
$^{\rm c}$ References for column density measurements. -- 
(1)   This work; 
(2)   H\'ebrard et al. (\citealp{hebrard99}); 
(3)   Kruk et al. (\citealp{kruk02}); 
(4)   Lemoine et al. (\citealp{lemoine02}); 
(5)   Wood et al. (\citealp{wood03}); 
(6)   Lehner et al. (\citealp{lehner02}); 
(7)   H\'ebrard et al. (\citealp{hebrard02}); 
(8)   Wood et al. (\citealp{wood02}); 
(9)   Oliveira et al. (\citealp{oliveira03}); 
(10)  York \& Kinahan (\citealp{york79});
(11)  Jenkins et al. (\citealp{jenkins99});
(12)  Meyer et al. (\citealp{meyer98}) \& Meyer~(\citealp{meyer01}); 
(13)  Ferlet et al. (\citealp{ferlet80});
(14)  Meyer et al. (\citealp{meyer97}); 
(15)  Laurent et al. (\citealp{laurent79}); 
(16)  Hibbert et al. (\citealp{hibbert85}); 
(17)  Friedman et al. (\citealp{friedman02}); 
(18)  Hoopes et al. (\citealp{hoopes03}) \\
\end{small}
\end{table*}

\begin{table}
\caption{Deuterium abundances.} 
\begin{tabular}{lcl}
\hline
D/O in the Local Bubble 
   & $3.84\pm0.16 $&$ \times 10^{-2}$ \\
D/O total in the Local Bubble (\ie\ corrected for oxygen depletion)
   & $\sim 2.7  $&$ \times 10^{-2}$ \\
D/O total expected in the local interstellar medium (LISM) 
from chemical evolution models 
   & $2.5 - 4 $&$ \times 10^{-2}$ \\
D/O at greater distances (\ie\ beyond the LISM) 
   & $1.50\pm0.25 $&$  \times 10^{-2}$ \\
D/O in QSO absorption lines 
   & $20 - 3000 $&$  \times 10^{-2}$ \\
\hline
D/H in the Local Bubble from previous direct measurements of 
$N$(\ion{D}{1}) and $N$(\ion{H}{1})
   & $1.5\pm0.1 $&$  \times 10^{-5}$ \\
D/H in the Local Bubble from D/O and O/H measurements 
   & $1.32\pm0.08 $&$  \times 10^{-5}$ \\
D/H at greater distances (\ie\ beyond the LISM) from D/O and O/H  
   & $0.52\pm0.09  $&$ \times 10^{-5}$ \\
D/H at greater distances (\ie\ beyond the LISM) from D/N and N/H  
   & $0.86\pm0.13  $&$ \times 10^{-5}$ \\
D/H in proto-solar material (Solar wind, Jupiter, Saturn)
   & $2-3  $&$ \times 10^{-5}$ \\
D/H in primordial material (QSO absorption lines and CMB) 
   & $2.5-3  $&$ \times 10^{-5}$ \\
\hline
\end{tabular}
\label{table_sum}
\end{table}

\end{document}